\documentclass[usenatbib]{mn2e}
\usepackage{graphicx,times, amsmath, epsfig, color}

\newcommand{\cmfast}{\textsc{\small 21CMFAST}}
\newcommand{\cmmc}{\textsc{\small 21CMMC}}
\newcommand{\muKK}{\mu {\rm K^2}}
\newcommand{\PkSZ}{[\Delta^{\rm patchy}_{l3000}]^2}

\newcommand{\avenf}{\bar{x}_{\rm HI}}

\newcommand{\lya}{Ly$\alpha$}
\newcommand{\lyb}{Ly$\beta$}

\newcommand{\Tvir}{T_{\rm vir}}
\newcommand{\Tvirmin}{T_{\rm vir}^{\rm min}}

\newcommand{\mfp}{R_{\rm mfp}}

\newcommand\lsim{\mathrel{\rlap{\lower4pt\hbox{\hskip1pt$\sim$}}
        \raise1pt\hbox{$<$}}}
\newcommand\gsim{\mathrel{\rlap{\lower4pt\hbox{\hskip1pt$\sim$}}
        \raise1pt\hbox{$>$}}}
\def\myputfigure#1#2#3#4#5%
{\vskip#5pt\makebox[0pt]{\hskip#2in
\includegraphics[width=#3\textwidth]{#1}}\vskip#4pt\hfill}

\newenvironment{packed_item}{
\begin{itemize}
  \setlength{\itemsep}{1pt}
  \setlength{\parskip}{0pt}
  \setlength{\parsep}{0pt}
}{\end{itemize}}

\newcommand{\hii}{\ifmmode\mathrm{H\,{\scriptscriptstyle II}}\else{}H\,{\scriptsize II}\fi}
\newcommand{\hi}{\ifmmode\mathrm{H\,{\scriptscriptstyle I}}\else{}H\,{\scriptsize I}\fi}

\pdfoutput=1

\begin{document}

\title[The global history of reionisation]{The Global History of Reionisation}

\author[Greig \& Mesinger]{Bradley Greig\thanks{email: bradley.greig@sns.it} \& Andrei Mesinger\\
Scuola Normale Superiore, Piazza dei Cavalieri 7, 56126 Pisa, Italy
}

\voffset-.8in

\maketitle

\begin{abstract}
Using a Bayesian framework, we quantify what current observations imply about the history of the epoch of reionisation (EoR).  We use a popular, three-parameter EoR model, flexible enough to accommodate a wide range of physically-plausible reionisation histories. We study the impact of various EoR observations: (i) the optical depth to the CMB measured by {\it Planck} 2016; (ii) the dark fraction in the Lyman $\alpha$ and $\beta$ forests; (iii) the redshift evolution of galactic \lya\ emission (so-called ``\lya\ fraction''); (iv) the clustering of \lya\ emitters; (v) the IGM damping wing imprint in the spectrum of QSO ULASJ1120+0641; (vi) and the patchy kinetic Sunyaev-Zel'dovich signal. Combined, (i) and (ii) already place interesting constraints on the reionisation history, with the epochs corresponding to an average neutral fraction of (75, 50, 25) per cent, constrained at 1$\sigma$ to $z= (9.21\substack{+1.22 \\ -1.15}, 8.14\substack{+1.08 \\ -1.00}, 7.26\substack{+1.13 \\ -0.96})$. Folding-in more model-dependent EoR observations [(iii--vi)], strengthens these constraints by tens of per cent, at the cost of a decrease in the likelihood of the best-fit model, driven mostly by (iii). The tightest constraints come from (v). Unfortunately, no current observational set is sufficient to break degeneracies and constrain the astrophysical EoR parameters. However, model-dependent priors on the EoR parameters themselves can be used to set tight limits by excluding regions of parameter space with strong degeneracies. Motivated by recent observations of $z\sim7$ faint, lensed galaxies, we show how a conservative upper limit on the virial temperature of haloes which host reionising galaxies can constrain the escape fraction of ionising photons to $f_{\rm esc} = 0.14\substack{+0.26 \\ -0.09}$.
\end{abstract}


\section{Introduction}
\label{sec:intro}

As the last major phase change of our Universe, the epoch of reionisation (EoR) encodes a wealth of information about the properties of the first structures.  Recent years have witnessed rapid progress in both observational evidence of this elusive epoch, as well as the theoretical tools required to interpret them.  We now have observational hints of the EoR from: (i) the electron scattering optical depth to the cosmic microwave background (CMB; e.g. \citealt{Komatsu11, Planck16}); (ii) the rapid redshift evolution of the mean transmission in the Lyman forests of $z\gsim6$ QSO spectra (e.g. \citealt{Fan01, White03}); (iii) the imprint of the intergalactic medium (IGM) damping wing in high-$z$ QSO spectra (e.g. \citealt{MH04, Carilli10, Mortlock11, Greig16}); (iv) the rapid disappearance of \lya\ emitting galaxies at $z\gsim6$ (e.g. \citealt{Stark10, Pentericci11, Ono12, Caruana14, Schenker14}).  Other observations, while not claiming to detect the EoR, nevertheless provide useful constraints.  These include: (i) the clustering of Lyman alpha emitters (LAEs; e.g. \citealt{Ouchi10}); (ii) the secondary anisotropies in the CMB (e.g. \citealt{George15}); (iii) the dark fraction of pixels in QSO spectra \citep{MMO15}; (iv) the distribution of dark gaps in QSO spectra (e.g. \citealt{Croft98, GCF06}); (v) upper limits on the cosmological 21-cm power spectrum (e.g. \citealt{Parsons14, Ali15}).

Although relatively numerous, these observations have been tricky to interpret, and consensus on what they imply about the EoR has been rare.  Nevertheless, interpretation has also been improving in recent years.  The theoretical front has been evolving from simple (yet highly inaccurate) models of reionisation as a homogeneous process in a uniform medium to the now ubiquitous models of inhomogeneous reionisation sourced by increasingly sophisticated galaxy formation prescriptions.  With this also came the realisation that we need to statistically account for the many inherent astrophysical uncertainties.  Efforts to statistically quantify EoR parameter constraints thus prompted the development of efficient, ``semi-numerical'' simulations (e.g. \citealt{Zahn07, MF07}), sub-grid modelling of missing physics (e.g. \citealt{Ciardi06, SM14}), tiered modelling involving hydrodynamic simulations ``nested'' inside large-scale semi-numerical models (e.g. \citealt{Mesinger15, Choudhury15}), ensemble averaging over medium size numerical simulations (e.g. \citealt{Sirko05, Gnedin14}), a variety of cosmic radiative transfer algorithms (e.g. \citealt{Iliev06, TG11}), and astrophysical parameter exploration using analytic/parametric models (e.g. \citealt{CF06, MCF15, Bouwens15, Khaire16, Mitra16, Price16}) as well as 3D EoR simulations (e.g. \citealt{McQuinn07, GM15}).

Such continuing observational and theoretical advances foreshadow that EoR studies will soon transition from an observationally-starved to a ``Big Data'' regime. These trends are set to peak with the advent of {\it 21cm tomography} with second generation interferometers (e.g. \citealt{Pober14, Koopmans15}).  The EoR morphology (i.e. the distribution of cosmic ionised and neutral patches) and its redshift evolution encode a wealth of information about the properties of galaxies and IGM structures (e.g. \citealt{McQuinn07, GM15}).

While preparing for this treasure trove of data, it is nevertheless instructive to ask what we can learn from current data, using the latest analysis techniques. The obvious starting point is the global history of the EoR, characterised by the redshift evolution of the mean IGM neutral hydrogen fraction, $\avenf$.  Even this simple statistic tells us when our galactic ancestors first appeared and how efficient they were at star formation.

In this paper, we apply the latest analysis frameworks to the most up to date EoR observations, showing the resulting constraints on the history of reionisation. Similar studies on the reionisation history have been performed previously, focusing on one or more observational constraints and using a variety of EoR models (e.g. \citealt{CF06, MCF11, KF-G12, Harker12, Zahn12, MMS12, Patil14, Bouwens15, MCF15, Khaire16, Mitra16, Price16}). Our work is unique due to a combination of the following reasons: (i) we make use of the latest EoR observations, presenting the corresponding constraints individually so as to separately highlight the impact of each observable; (ii) we take constraints on $\avenf$ directly from the latest, most sophisticated investigations, tailored to each  observable; (iii)  our EoR model is flexible enough to accommodate a wide range of physically-motivated reionisation models; (iv) we use a Bayesian framework which samples {\it 3D simulations}, allowing us to efficiently characterise signatures which depend on the EoR spatial structure (such as the kinetic Sunyaev-Zel'dovich signal, as well as the late-time, sink-dominated evolution of $\avenf$).

The remainder of this paper is organised as follows. In Section~\ref{sec:model}, we summarise our analysis framework. In Section~\ref{sec:gold} we introduce our ``Gold Sample'', consisting of observational constraints with little or no dependence on EoR modelling.  Next, in Section~\ref{sec:model_depend} we fold-in model-dependent EoR observations, showing their corresponding impact on the reionisation history and EoR parameter constraints, both individually and combined with the Gold Sample. We then combine all of our observational constraints in Section~\ref{sec:everything} and provide a short discussion on model-dependent observational priors on our EoR parameters in Section~\ref{sec:LFobs}.   Finally, we conclude in Section~\ref{sec:conc}. Unless stated otherwise, we quote all quantities in comoving units. We adopt the background cosmological parameters: ($\Omega_\Lambda$, $\Omega_{\rm M}$, $\Omega_b$, $n$, $\sigma_8$, $H_0$) = (0.69, 0.31, 0.048, 0.97, 0.81, 68 km s$^{-1}$ Mpc$^{-1}$), consistent with recent results from the Planck mission \citep{Planck15}.

\section{Analysis framework}
\label{sec:model}

\subsection{Modelling reionisation}
Our analysis uses Bayesian sampling of the semi-numerical EoR simulation code \cmfast\footnote{http://homepage.sns.it/mesinger/Sim} \citep{MF07, MFC11}.  Our fiducial simulations are $L$ = 250~Mpc on a side, initialised with the 2015 Planck $\Lambda$CDM cosmological parameters \citep{Planck15}, and down-sampled to a final resolution of 256$^3$. For a given EoR parameter set, the ionisation field is constructed by comparing the number of ionising photons to the number of baryons (plus recombinations), following an excursion-set approach for patchy reionisation \citep{FHZ04}.  Specifically, a simulation cell is flagged as ionised if $\zeta f_{\rm coll}(\boldsymbol{x},z,R,\Tvirmin) \geq 1$, where $\zeta$ is the ionisation efficiency, and $f_{\rm coll}$ is the fraction of matter inside a region of size $R$ residing within haloes with virial temperatures larger than some threshold required for efficient star-formation, $\Tvirmin$. 
The sampling of the EoR parameters (listed below) is performed with the Markov Chain Monte Carlo (MCMC) framework, \cmmc\footnote{https://github.com/BradGreig/21CMMC} \citep{GM15}, modified to sample from a fixed, high-resolution grid\footnote{Here we repeatedly perform the MCMC sampling for various combinations of observational priors, but using an EoR model with only three parameters.  Moreover, current measurements only indirectly constrain the global properties of the reionisation epoch (e.g. $\avenf$, $\tau_{\rm e}$), whose values vary slowly and smoothly in our three parameter EoR model (i.e. without sharp peaks in the likelihood surface). Hence, for computational efficiency, we  replace the on-the-fly generation of 3D EoR boxes in each step in the chain with interpolation from a high-resolution fixed grid.  Using one representative observational prior, we confirm that this modification has no noticeable impact on the derived constraints by comparing the results to those obtained with the unmodified \cmmc\ sampler.}. For further details on the model, interested readers are encouraged to read the above-cited articles.

We sample three fundamental EoR parameters:
\begin{packed_item}
\item {\bf Minimum virial temperature of star-forming galaxies, $\boldsymbol{\Tvirmin}$} --  The minimum virial temperature of star-forming galaxies is set by the requirement for gas to condense and cool inside dark matter haloes, and form stars efficiently in the presence of SNe and photo-heating feedback mechanisms. $T^{\rm min}_{\rm vir}$ affects (i) when reionisation occurs, and (ii) the bias of the galaxies driving it.  A higher value of $T^{\rm min}_{\rm vir}$ means that reionisation happened later, with more large-scale ionisation structure (at a fixed value of the mean neutral fraction; e.g. \citealt{McQuinn07}). Here we assume a flat prior over the log of the virial temperature, within the range  $T^{\rm min}_{\rm vir}\in[10^{4},5\times10^{5}]$~K\footnote{
Note that this lower limit of $T^{\rm min}_{\rm vir} = 10^4$ K is roughly consistent with the fiducial choices in  \citet{Robertson13} and \citet{Robertson15}. Here, these authors determine that in order to achieve reionisation by $z\sim6$, they must extrapolate the observed luminosity function down to (at least) $M_{\rm UV} = -13$. This limit, under the scaling provided in Equation~\ref{eq:conversion} which assumes a duty cycle of 0.5, corresponds to $T^{\rm min}_{\rm vir} \sim 2\times10^4$ K.}
.  The lower limit corresponds to the atomic cooling threshold and the upper limit is roughly consistent with the host haloes of observed Lyman break galaxies at $z\sim6$--8 (e.g. \citealt{KF-G12, Barone-Nugent14}).  For reference, the virial temperature can be related to the halo mass via, (e.g. \citealt{BL01}):
\begin{eqnarray}
  M_{\rm halo}^{\rm min} &=& 10^{8} h^{-1} \left(\frac{\mu}{0.6}\right)^{-3/2}\left(\frac{\Omega_{\rm m}}{\Omega^{z}_{\rm m}}
\frac{\Delta_{\rm c}}{18\pi^{2}}\right)^{-1/2} \nonumber \\
& & \times \left(\frac{T^{\rm min}_{\rm vir}}{1.98\times10^{4}~{\rm K}}\right)^{3/2}\left(\frac{1+z}{10}\right)^{-3/2}M_{\sun},
\end{eqnarray}
where $\mu$ is the mean molecular weight, $\Omega^{z}_{\rm m} = \Omega_{\rm m}(1+z)^{3}/[\Omega_{\rm m}(1+z)^{3} + 
\Omega_{\Lambda}]$, and $\Delta_{c} = 18\pi^{2} + 82d - 39d^{2}$ where $d = \Omega^{z}_{\rm m}-1$. 

In reality, there will be broad scatter in the efficiency of star formation in haloes around $\Tvirmin$, resulting in a flattening of the ionising luminosity function instead of a sharp drop (e.g. \citealt{Liu16}).
However, the progress of reionisation only depends on the integral over the ionising luminosity function, which is dominated by average properties and abundances of the faint galaxies due to the steepness of the halo mass function and the likelihood that small galaxies have higher ionising escape fractions (e.g. \citealt{PKD15}).  Thus, regardless of the precise shape of the ionising luminosity function at the faint end, our $\Tvirmin$ parameter provides a straightforward {\it proxy for the typical haloes hosting ionising sources during reionisation.},

\item {\bf Galactic ionising efficiency, $\boldsymbol{\zeta}$} -- Galaxies hosted in haloes with virial temperatures greater than $\Tvirmin$ are assumed to have ionising efficiencies of $\zeta = 30\left(\frac{f_{\rm esc}}{0.2}\right)\left(\frac{f_{\ast}}{0.03}\right) \left(\frac{f_b}{\Omega_b/\Omega_m}\right)
  \left(\frac{N_{\gamma/b}}{4000}\right)
  \left(\frac{1.5}{1+n_{\rm rec}}\right)$,
  where $f_{\rm esc}$ is the fraction of ionising photons escaping into the IGM,
  $f_b$ is the baryon fraction inside haloes hosting galaxies in units of the cosmic baryon fraction,
  $f_\ast$ is the fraction of galactic gas in stars, $N_{\gamma/b}$ is the number of ionising photons per baryon in stars and $n_{\rm rec}$ is the average number of recombinations per baryon in the IGM. The parameter $\zeta$ mainly serves to speed-up/slow-down reionisation.  Within this work, we take a flat prior over the range $\zeta\in[5,200]$, which results in a range of reionisation histories which are in broad agreement with current EoR constraints as we shall see below.

\item {\bf Ionising photon horizon through the ionised IGM, $\boldsymbol{\mfp}$} -- ionising photons escaping galaxies can be re-absorbed by recombinations inside their local cosmic \hii{} patch.  When the typical distance ionising photons can travel through the ionised IGM is smaller than the typical \hii{} region, an increasing number of photons are lost to recombinations and reionisation slows down (e.g. \citealt{FO05, FM09, AA12}).  This maximum horizon for ionising photons (commonly referred to as a ``mean free path'' following its instantaneous, Str{\"o}mgren sphere limit) is determined by the requirement that the time-integrated number of ionising photons in a given region is equal to or greater than the number of baryons plus the time-integrated number of recombinations (e.g. \citealt{SM14}). 
Here we implement $\mfp$ as the maximum filtering scale in our reionisation algorithm, and adopt a flat prior over the range $\mfp\in[5, 40]$~Mpc, motivated by sub-grid models of inhomogeneous recombinations \citep{SM14}, as well analytic estimates \citep{FO05} and hydrodynamic simulations of the IGM \citep{MOF11, ETA13}.
\end{packed_item}

\begin{figure}
{
\includegraphics[trim = 0.8cm 1.8cm 0cm 0cm, scale = 0.45]{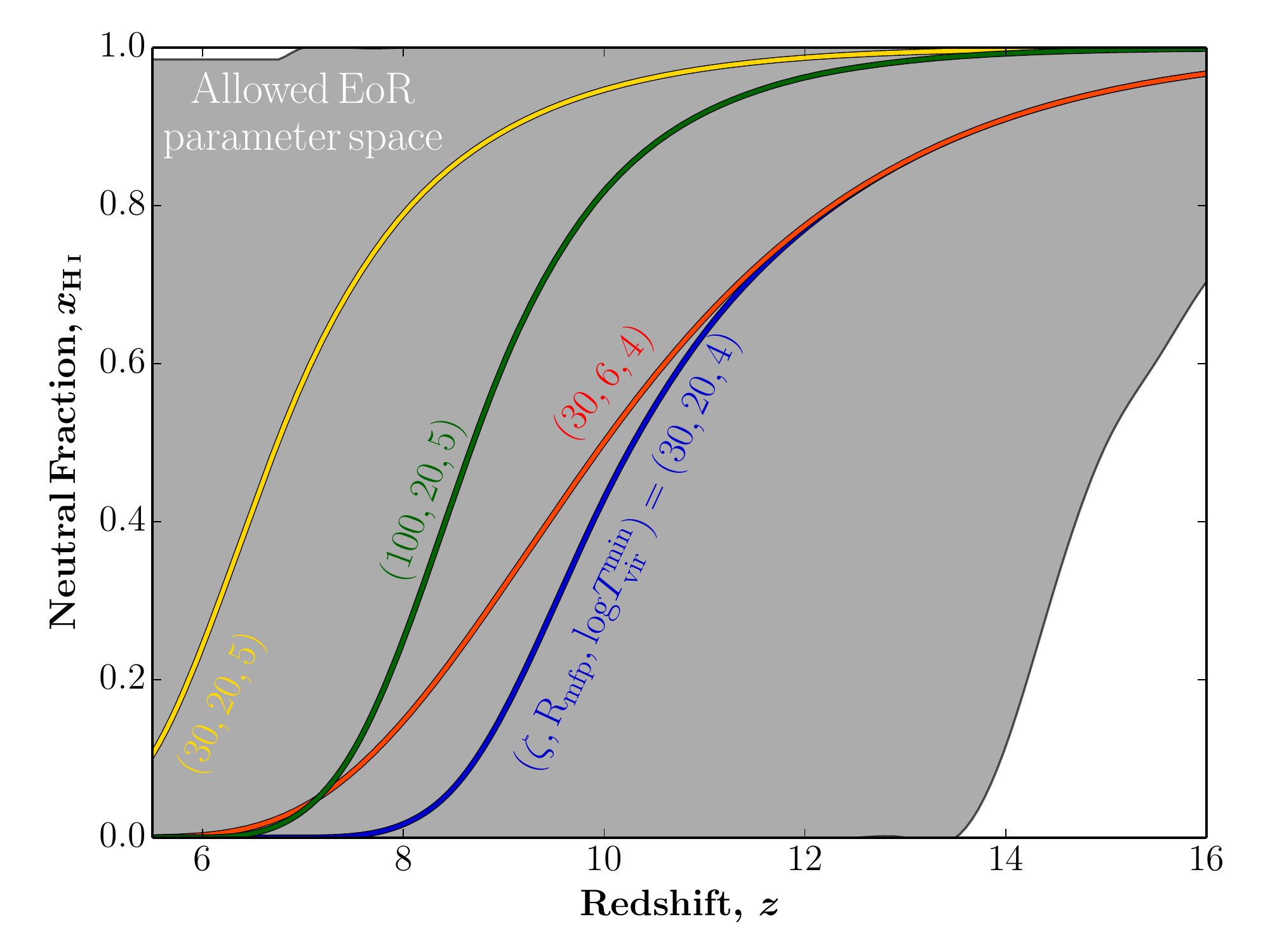}
}
\caption{
The range of reionisation histories sampled by our three parameter EoR model.  We also show four individual histories, to illustrate the impact of each model parameter.
}
\label{fig:sampling}
\vspace{-0.5\baselineskip}
\end{figure}

Even though this three-parameter EoR model is overly simplistic, it is well-suited for our purposes for several reasons. Although they average over stochasticity as well as redshift and halo mass dependencies in the source populations, these ``effective'' EoR parameters have a straightforward physical interpretation.  This is in contrast with the popular, unphysical parametrisation of the reionisation history (for example, using $\tanh(z)$).  Having a relatively-obvious physical meaning allows us to learn about galaxy formation from the reionisation history.
This connection to the underlying EoR astrophysics is also more direct than in models using 
empirical parameters based on Lyman break galaxy candidates (LBGs; e.g. limiting magnitudes, star formation histories, etc.).  This is because the dominant ionising population most likely consists of galaxies much fainter than current detection limits (e.g. \citealt{CFG08, KF-G12, YFX16}) thus requiring uncertain extrapolations both in magnitude and redshift.  Moreover, $f_{\rm esc}$ is unknown for even the observed LBGs, let alone for the unobserved population likely driving reionisation.

Finally and  most importantly,  this EoR parametrisation is {\it flexible}, allowing for a large range of EoR histories which might be unfairly excluded by more sophisticated (but inaccurate) models.  The last point is illustrated explicitly in Fig. \ref{fig:sampling}, where we show the range of EoR histories sampled by our EoR parameters.  This range easily accommodates all physically-motivated EoR histories.

Although this three parameter model serves primarily to provide a flexible basis set of EoR histories, it is useful to speculate how these can be related to more sophisticated EoR physics.  As an example, photo-heating feedback \citep[e.g.][]{Iliev07,MD08} and inhomogeneous recombinations \citep{SM14} can delay reionisation, especially in the final stages.  The EoR histories of these models can be roughly mimicked by adopting a low value for $R_{\rm mfp}$ (Mesinger et al., in prep). It is more challenging for this three parameter model to mimic more extreme EoR histories which include a low-ionisation tail towards high redshifts, motivated by very efficient minihalo star formation \citep{Ahn12} or very efficient X-ray ionisation \citep{RO4, MFS13}.  However, most current observations are not very sensitive to such EoR histories, and testing such models will likely have to be done with upcoming 21-cm observations.  The exception to this is the Planck 2016 measurement of $\tau_{e}$, which limits the allowed parameter space for such extreme models.  In the context of our three-parameter model, an extended low-ionization tail towards high-$z$ would be similar to taking an even lower effective $\tau_{e}$ for the remainder of the EoR, pushing the bulk of reionization even later in order to compensate for an earlier start.

\subsection{Casting into empirical galaxy parameters, $M_{\rm UV}$ and $f_{\rm esc}$}
\label{sec:empirical}

\begin{figure}
{
\includegraphics[trim = 0.4cm 1cm 0cm 0cm, scale = 0.46]{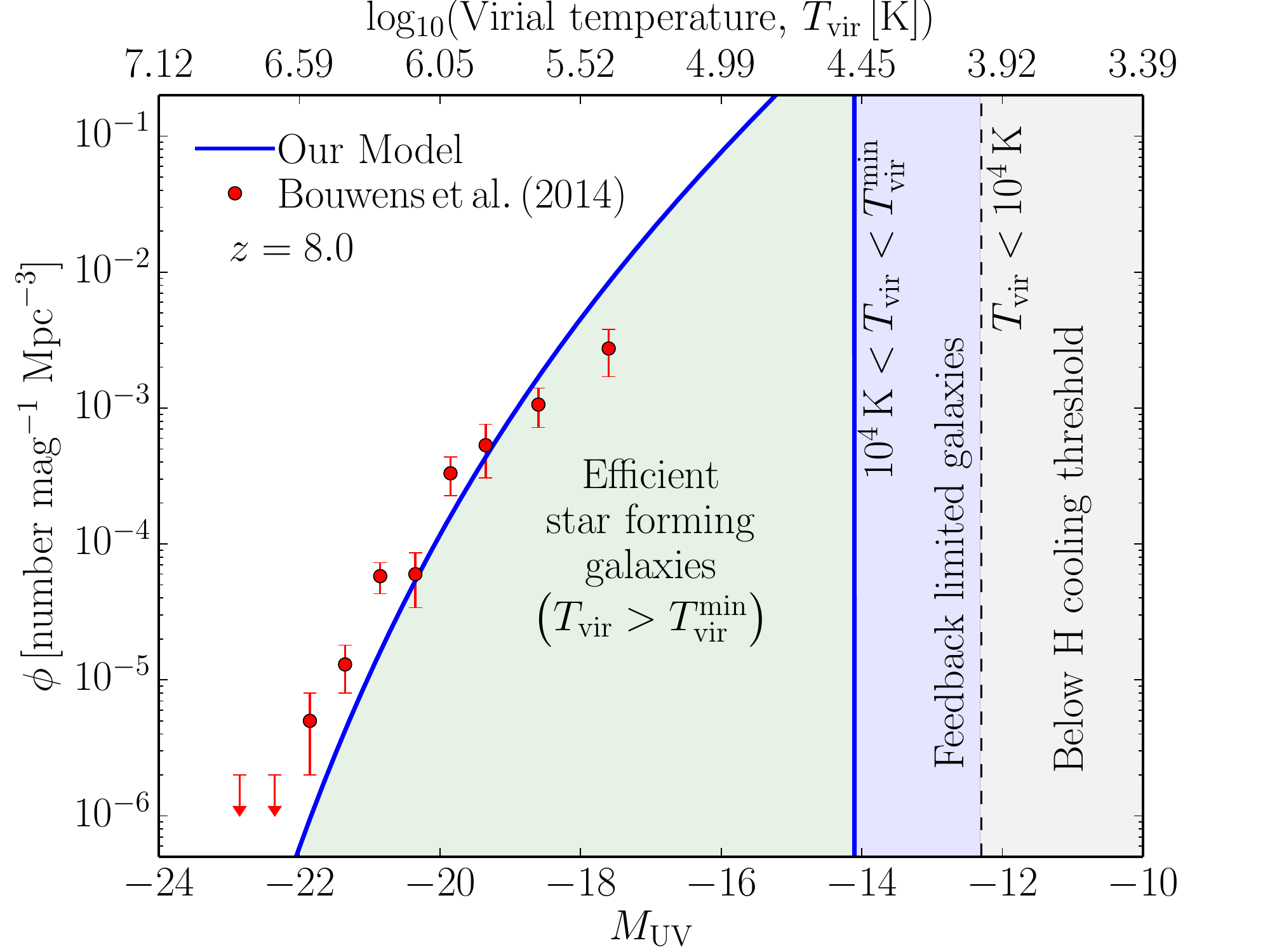}
}
\caption{
A schematic figure showing how our theoretical source parameters can map onto empirical LBG parameters.
Galaxies inside haloes with $T_{\rm vir} > \Tvirmin$ can efficiently form stars, while star formation inside smaller haloes is inhibited by feedback mechanisms.
The relation between the virial temperature ({\it top axis}) and the non-ionising UV magnitude ({\it bottom axis}) is obtained through abundance matching, assuming a duty cycle of 0.5.  The blue curve corresponds to a fiducial scaling of $L_{1500} \propto M_{\rm halo}$ (see text for details).  Note that a very sharp drop at $T_{\rm vir} < \Tvirmin$ (or $M_{\rm UV}> M_{\rm UV}^{\rm min}$) is unlikely, due to stochasticity in the star formation efficiency of low mass galaxies.  However, irrespective of the precise shape of the luminosity function, our $\Tvirmin$ parameter is a proxy for the typical haloes hosting the dominant ionising population during the EoR.
}
\label{fig:LBGs}
\vspace{-0.5\baselineskip}
\end{figure}

As mentioned above, our physically-motivated model provides a more direct parametrisation of the EoR than those based on empirical galaxy parameters.  It is nevertheless useful to relate these parametrisations.  We illustrate one such mapping in Fig. \ref{fig:LBGs}.  At reionisation redshifts, the relation between the virial temperature ({\it top axis}) and the non-ionising UV magnitude ({\it bottom axis}) can be roughly expressed as:
\begin{eqnarray} \label{eq:conversion}
{\rm log}_{10}(T_{\rm vir}/{\rm K}) \simeq {\rm log}_{10}(5.23) - \frac{4}{15}M_{\rm UV}.
\end{eqnarray}
To obtain this expression, we begin with the UV luminosity function as a function of UV magnitude, $\phi(M_{\rm UV}) = \frac{dn}{dM_{\rm halo}}\line(0,1){10}_{\line(0,1){10}}\frac{dM_{\rm halo}}{dM_{\rm UV}}\line(0,1){10}_{\line(0,1){10}}$. For simplicity, we then assume a linear proportionality between the halo mass and the 1500\AA\ luminosity ($M_{\rm halo}\propto L_{1500}$) and convert this to a function of $M_{\rm UV}$ using the typical AB magnitude relation, ${\rm log_{10}}(L_{1500}/({\rm erg\,s^{-1}\,Hz^{-1}})) = 0.4(51.63-M_{\rm UV})$. The constant of proportionality for this linear scaling is then determined directly from abundance matching (which determines the expression for $\frac{dM_{\rm halo}}{dM_{\rm UV}}$), assuming a duty cycle of 50 per cent\footnote{Note, the normalisation of this expression was not determined in order to best fit the observational data in Figure~\ref{fig:LBGs}. Rather, it was selected to provide a reasonable match to the observational data while at the same time facilitating the simple functional form presented in Equation~\ref{eq:scaling}.}. Following this, we recover $M_{\rm halo} = 1331.88\times10^{-0.4 M_{\rm UV}}M_{\odot}$ with which we then convert the expression for the virial temperature ($T_{\rm vir}$) as a function of halo mass (equation 26 of \citealt{BL01}) into a function of UV magnitude at $z=8$.

These choices are partially degenerate with the scalings of the additional parameters below.  In particular, while a linear scaling does not provide the best fit to the $z\approx8$ LFs, it facilitates a direct conversion from our fiducial EoR parameters, and suffices for a rough estimate.

To relate the ionising efficiency, $\zeta$, to the ionising photon escape fraction, $f_{\rm esc}$, we assume that the escape fraction and dust obscuration are constant.  In this framework, the emission rate of ionising photons for a galaxy residing in a halo of mass $M_{\rm halo}$ can be expressed as\footnote{As discussed above, the relevant parameter in our model is the {\it time-integrated} number of ionising photons.  For simplicity of conversion, here we relate the time-integrated and instantaneous emission rates with a characteristic star-formation time-scale, $t_\ast$.}: 
$\dot{N}_{\rm ion} \approx f_{\rm esc} f_\ast N_{\gamma/b} M_{\rm halo} f_b \Omega_{\rm b}\Omega^{-1}_{\rm m} m^{-1}_{\rm H} t^{-1}_\ast = \zeta (1+n_{\rm rec}) \Omega_{\rm b} \Omega^{-1}_{\rm m}m_{\rm H}^{-1} t_\ast^{-1} M_{\rm halo}$. We can relate this expression to the corresponding UV luminosity, by adopting $\dot{N}_{\rm ion} \propto f_{\rm esc} \gamma_{\rm ion}(L_{1500})$, where $\gamma_{\rm ion} = 2\times10^{25}~{\rm s^{-1}}\left[L_{\rm 1500}/({\rm erg ~ s^{-1} Hz^{-1}}) \right]$ relates the UV luminosity to the ionising luminosity, using a fiducial spectral energy density (SED) profile \citep[e.g.][]{KF-G12}.
Equating these expressions, and using the $M_{\rm halo} \propto L_{1500}$ normalisation from above (corresponding to the blue curve in Fig. \ref{fig:LBGs}) results in:
\begin{eqnarray} \label{eq:scaling}
\zeta = 200\left(\frac{\Omega_{\rm m}}{0.308}\frac{0.048}{\Omega_{\rm b}}\right)\left(\frac{1.5}{1+n_{\rm rec}}\right)\left(\frac{t_{\ast}}{250~{\rm Myr}}\right)f_{\rm esc}.
\end{eqnarray}

We use these relations to crudely map our fiducial EoR parameters, $\Tvirmin$ and $\zeta$, to popular LBG parameters, $M_{\rm UV}^{\rm min}$ and $f_{\rm esc}$, showing the latter on the top axis of our parameter constraints plots.
We stress however that the mappings above are not unique\footnote{We note that the ionising photon production during the bulk of the EoR is degenerate with redshift evolution in the other parameters.    For example, the emissivity in a `fiducial' model with an EoR midpoint at $z=7.5$ given by $\Tvir=3\times10^4$ K, $R_{\rm mfp} = 20$ Mpc, and a constant $\zeta$ = 20 can be reproduced by changing $\zeta = 11 \times \left[\Tvir/(3\times10^4 K)\right]^{0.45}$.
}, and depend on the assumed star formation histories, the halo mass and redshift scalings of the galactic SEDs, escape fraction, and duty cycle.

\subsection{Quantifying EoR constraints}

\begin{table*}
\begin{tabular}{@{}lccccccc}
\hline
Observational Priors & $z(\avenf=0.25)$ & $z(\avenf=0.5)$ & $z(\avenf=0.75)$ & $\chi^{2}_{\rm tot}$\\
\hline
\vspace{0.8mm}
Dark fraction & $10.27\substack{+3.37 \\ -3.74}$ &  $11.53\substack{+2.74 \\ -4.20}$ & $12.79\substack{+2.62 \\ -4.46}$ & 0.0\\
\vspace{0.8mm}
$\tau_e$ & $\lsim8.18$ & $7.64\substack{+1.34 \\ -1.82}$& $8.70\substack{+1.48 \\ -1.64}$ & 0.01 \\
\vspace{0.8mm}
Dark fraction + $\tau_e$ (Gold sample) & $7.26\substack{+1.13 \\ -0.96}$ & $8.14\substack{+1.08 \\ -1.00}$ & $9.21\substack{+1.22 \\ -1.15}$ & 0.08 \\
\hline
\lya\ fraction & $\lsim8.09$ & $\lsim9.25$ & $\lsim10.13$ & 1.18 \\
\vspace{0.8mm}
\lya\ fraction + Gold sample & $7.07\substack{+1.04 \\ -0.87}$ & $7.89\substack{+1.09 \\ -0.91}$ & $8.89\substack{+1.33 \\ -1.06}$ & 1.61\\
\hline
LAE clustering & $11.02\substack{+2.61 \\ -5.46}$ & $11.59\substack{+2.68 \\ -5.05}$ & $12.85\substack{+2.57 \\ -5.31}$ & 0.0 \\
\vspace{0.8mm}
LAE clustering + Gold sample & $7.32\substack{+1.09 \\ -0.98}$ & $8.13\substack{+1.11 \\ -0.96}$ & $9.21\substack{+1.27 \\ -1.10}$ & 0.13 \\
\hline
QSO damping wing & $6.57\substack{+0.68 \\ -0.89}$ & $7.32\substack{+0.76 \\ -0.73}$ & $8.39\substack{+0.85 \\ -0.77}$ & 0.01\\
\vspace{0.8mm}
QSO damping wing + Gold sample & $6.94\substack{+0.78 \\ -0.79}$ & $7.76\substack{+0.80 \\ -0.81}$ & $8.70\substack{+1.03 \\ -0.89}$ & 0.23 \\
\hline
kSZ & $\lsim9.29$ & $\lsim10.25$ & $\lsim11.32$ & 0.01\\
\vspace{0.8mm}
kSZ + Gold sample & $7.20\substack{+1.01 \\ -0.93}$ & $8.01\substack{+1.04 \\ -0.96}$ & $9.02\substack{+1.24 \\ -1.08}$ & 0.45\\
\hline
All priors combined & $6.82\substack{+0.78 \\ -0.71}$ & $7.57\substack{+0.78 \\ -0.73}$ & $8.52\substack{+0.96 \\ -0.87}$ & 2.15\\
\hline
\end{tabular}
\caption{Parameter constraints for each observational prior, and combinations of priors.  Values correspond to the peak likelihood of the individual marginalised 1D PDFs, with the quoted uncertainties enclosing 68 per cent of the total probability.  Entries which are only upper or lower limits correspond to values enclosing  68 per cent of the total probability (see text for details).}
\label{tab:Sum}
\end{table*} 

Throughout this work, we report the impact of the observational priors through constraints on both the reionisation history and on our three EoR model parameters. Within \cmmc\ we perform a MCMC maximum likelihood sampling\footnote{
We use a $\chi^2$ value as our maximum likelihood, determined by comparing the mock observed quantity to the actual observed constraint. For example, in the case of the electron scattering optical depth, $\tau_{e}$, $\chi^2 = (\tau_{e, {\rm obs}} - \tau_{e, {\rm mod}})^2/\sigma^2_{\tau_{e}}$, where $\tau_{e, {\rm mod}}$ is the modelled value determined from this three EoR parameter dataset and $\tau_{e, {\rm obs}} \pm \sigma_{\tau_{e}}$ from Planck \citep{Planck16}. 
} of our three EoR model parameters,  to recover both the 1D and 2D marginalised probability distribution functions (PDFs). 
For the 2D PDFs, the shaded regions correspond to the recovered 68 and 95 percentile (1 and 2$\sigma$) marginalised constraints.  Constraints on the reionisation history are obtained by constructing a 2D smoothed histogram of the $\avenf$-$z$ parameter space from the entire MCMC sampled data set recovered by \cmmc. The recovered 1 and 2$\sigma$ contours then enclose the 68 and 95 percentiles of the distribution in each redshift bin.

In Table~\ref{tab:Sum}, we provide a summary of the recovered constraints on the redshifts corresponding to 25, 50 and 75 per cent ionisation, for each observational prior, as well as the range enclosing 68 per cent of the total probability.  We also list the $\chi^{2}_{\rm tot}$ of the maximum likelihood model for each combination of observational constraints. To obtain $\chi^{2}_{\rm tot}$, we linearly add the priors for each corresponding observational constraint in log-likelihood space. Note that in a few instances we return a $\chi^{2}_{\rm tot} = 0$. This only arises in cases where the observed prior is an upper limit, for which we model as a one-sided Gaussian (see e.g. Section~\ref{sec:dark} and~\ref{sec:LAEclust}). Model parameter sets that meet this upper limit criterion are then assigned a probability of unity ($\chi^{2}_{\rm tot} = 0$).

\section{Model-independent EoR observational priors: the ``Gold sample''}
\label{sec:gold}

  We start with a ``gold sample'' of EoR constraints, consisting of those observations whose interpretation has little or no dependence on EoR modelling.  We then fold-in additional observational priors, one at a time, showing their impact on the reionisation history and EoR model parameters.

\subsection{The dark fraction in the Lyman alpha forest} \label{sec:dark}

\begin{figure*}
{
\includegraphics[trim = 1.1cm 0.7cm 0cm 0.3cm, scale = 0.445]{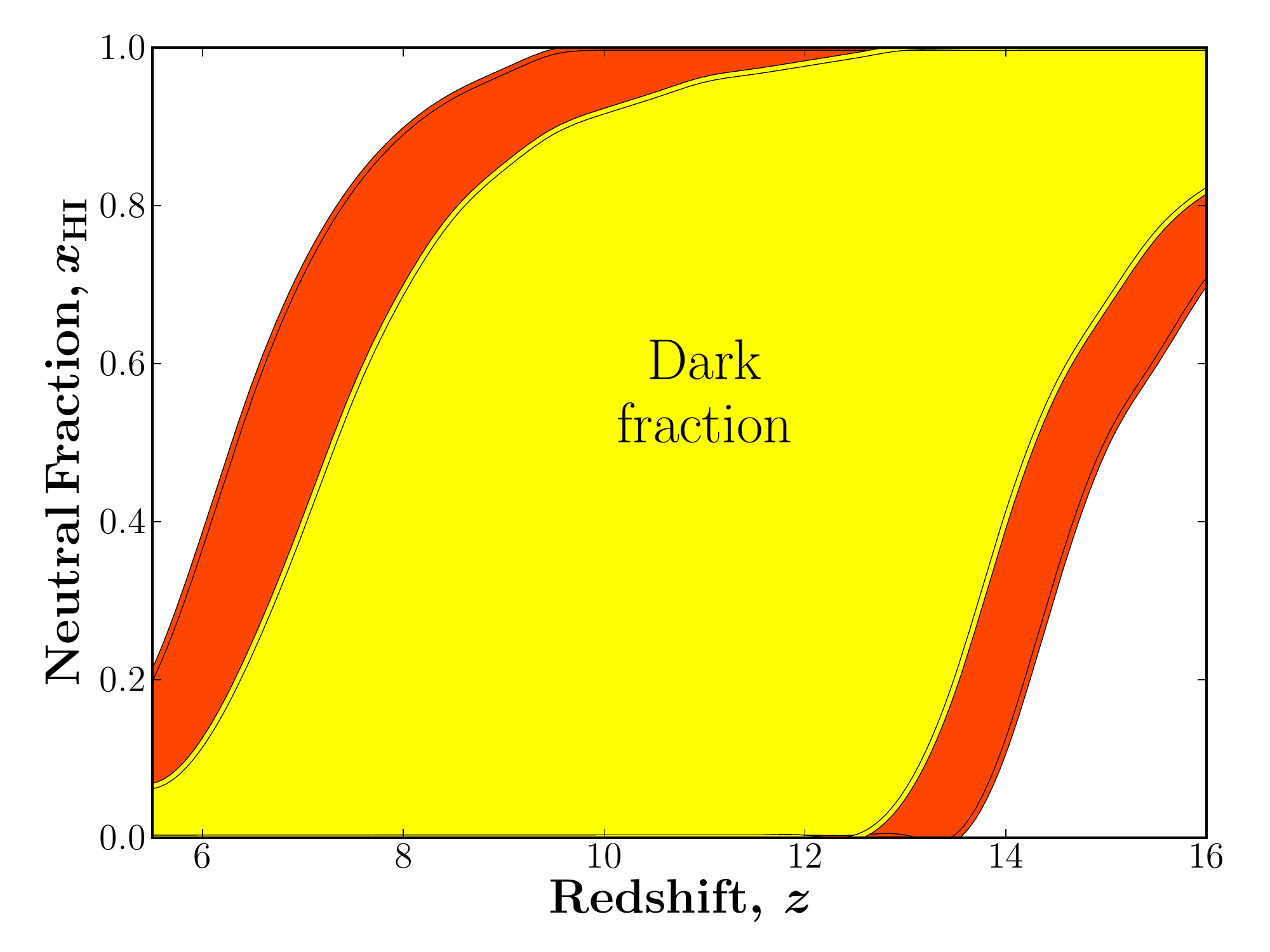}
\includegraphics[trim = 0.8cm 0.2cm 0cm 0.3cm, scale = 0.455]{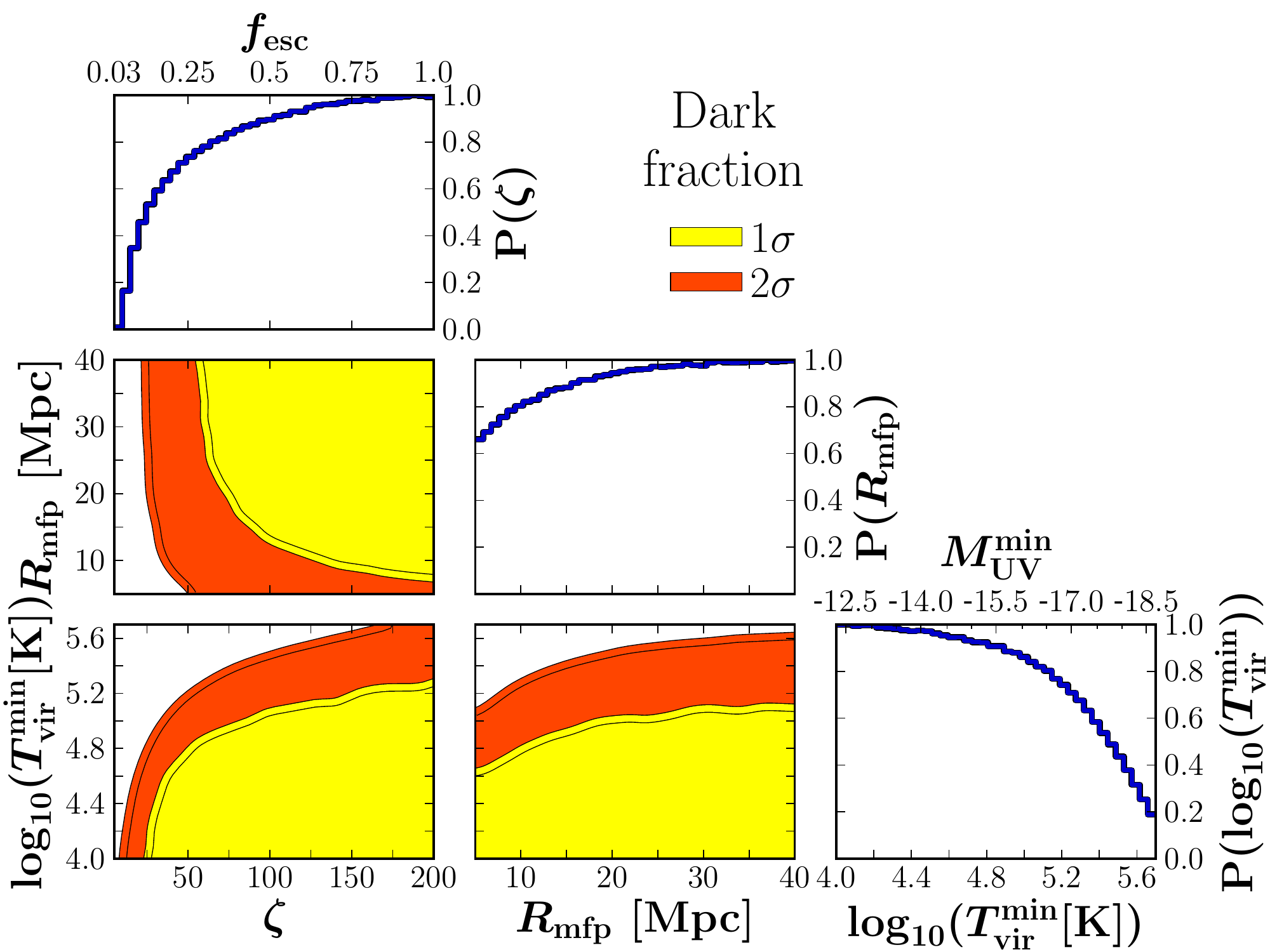}
}
\caption{
{\it Left:} the range of reionisation histories which are consistent with the dark fraction constraint of $\avenf(z=5.9) \lsim 0.11~(1\sigma)$ from \citet{MMO15}.
{\it Right:} constraints on our EoR model parameters implied by the dark fraction upper limit.  The diagonal panels show the 1D marginalised PDFs for each parameter: $\zeta$, $R_{\rm mfp}$ and  log$_{10}$$(\Tvirmin)$ ({\it from upper left to bottom right}).  The joint 2D marginalised likelihood contours are shown in the 3 panels in the bottom left.
In all panels, the 1$\sigma$ constraints are shown in yellow, and $2\sigma$ constraints are shown in red.
}
\label{fig:darkfraction}
\vspace{-0.5\baselineskip}
\end{figure*}

The only constraint on $\avenf$ completely independent of EoR modelling comes from the so-called ``dark fraction'' of QSO spectra \citep{Mesinger10}.  A zero-flux pixel in the Lyman $\alpha$ or $\beta$ forests of high-$z$ QSOs can result from either (i) a cosmic \hi{} patch (with a neutral fraction of $\sim$ unity); or (ii) the residual \hi{} (with a neutral fraction of $\gsim10^{-4}$) inside the ionised IGM.  Discriminating between (i) and (ii) requires knowledge of the bias of cosmic \hi{} patches during the EoR, as well as the density, ionising background and their cross-correlation inside the cosmic \hii{} patches.  If however one does not attempt to discriminate between these two possible sources of saturated pixels, one obtains a less stringent but model-independent upper limit on $\avenf$ simply from the fraction of pixels which are dark (with zero flux).  Such an upper limit can be powerful if constructed from high signal-to-noise spectra which resolve the bulk of the transmission inside the ionised IGM.

By noting the fraction of pixels which are dark in {\it both} the \lya\ and \lyb\ forests in a sample of 22 QSO spectra, \citet{MMO15} recently obtained upper limits on $\avenf$ in the redshift range $z\approx$ 5--6.  The most relevant of these (i.e. the tightest limit at the highest redshift) is the 1$\sigma$ upper limit of $\avenf \lsim 0.11$ at $z=5.9$.

In Fig. \ref{fig:darkfraction} we show the resulting constraints on the reionisation histories ({\it left}) and our three EoR model parameters ({\it right}).  We implement the $\avenf(z=5.9) \lsim 0.06 + 0.05 ~ (1\sigma)$ constraint by taking a flat prior over $\avenf < 0.06$, and at higher values a one-sided Gaussian with a $\sigma=0.05$.

In all panels, the 1$\sigma$ constraints are shown in yellow, and $2\sigma$ constraints are shown in red.  
By comparing the left panel with Fig. \ref{fig:sampling}, we see that the  dark fraction upper limit rules out the upper left corner of the reionisation history parameter space.  These models would have reionisation finish later than implied by the dark fraction constraints.  Conversely, the constraints at high redshift are unaffected by this prior on $\avenf(z=5.9)$. Limits on the the early stages $z>13$ are the result of the sampling of our parameter space, as can be seen comparing to Fig.~\ref{fig:sampling} which does not include any priors.

The impact of the dark fraction prior on the EoR parameters is seen in the right panels of Fig.~\ref{fig:darkfraction}.  The parameter space which results in late finishing reionisation is disfavoured, i.e. galaxies which are inefficient (low $\zeta$), late-forming and rare (high $\Tvirmin$), and which are surrounded by a rapidly recombining IGM  (small $\mfp$).

\subsection{The optical depth to the CMB}

\begin{figure*}
{
\includegraphics[trim = 1.1cm 0.7cm 0cm 0.3cm, scale = 0.445]{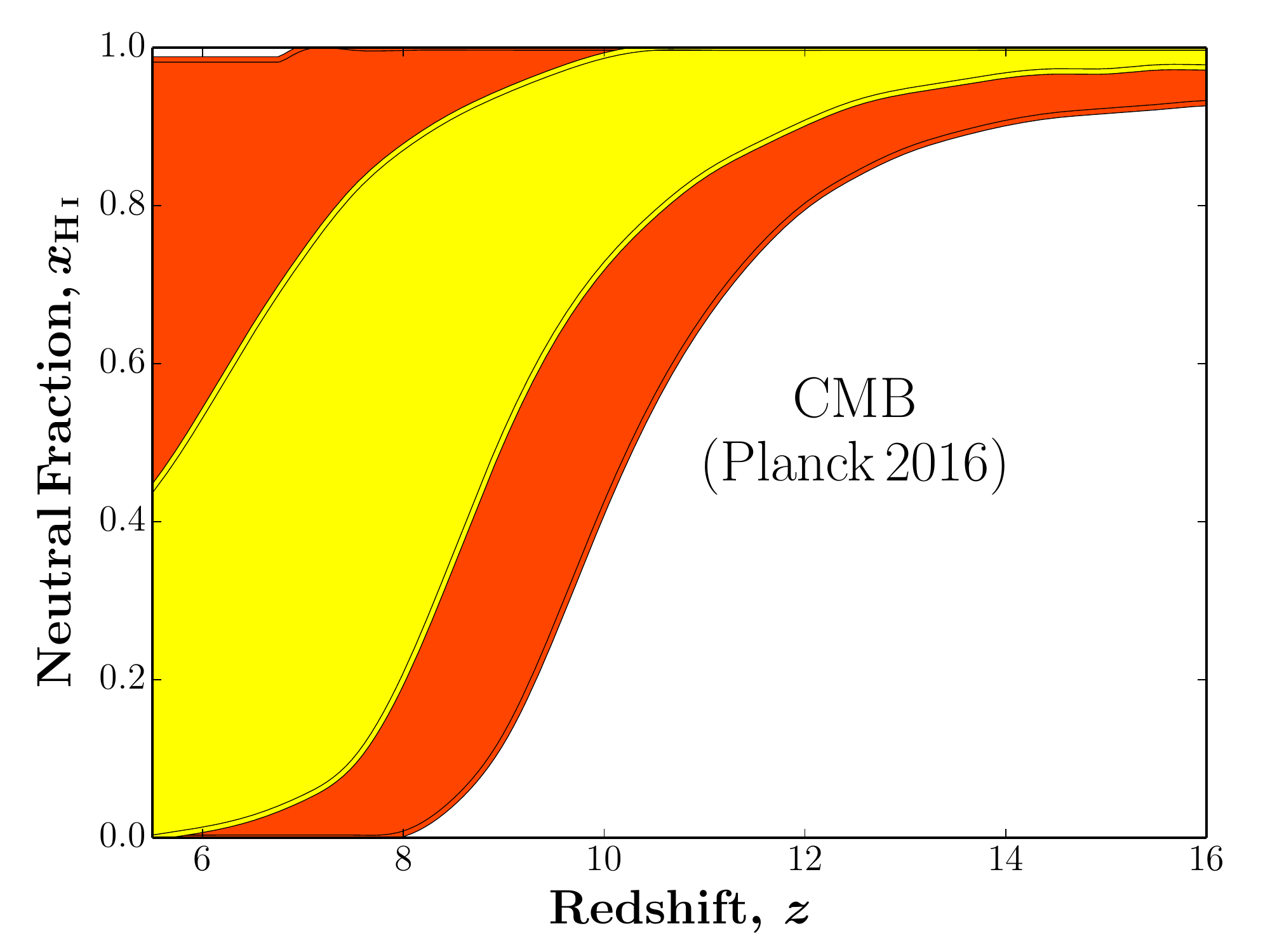}
\includegraphics[trim = 0.8cm 0.2cm 0cm 0.3cm, scale = 0.455]{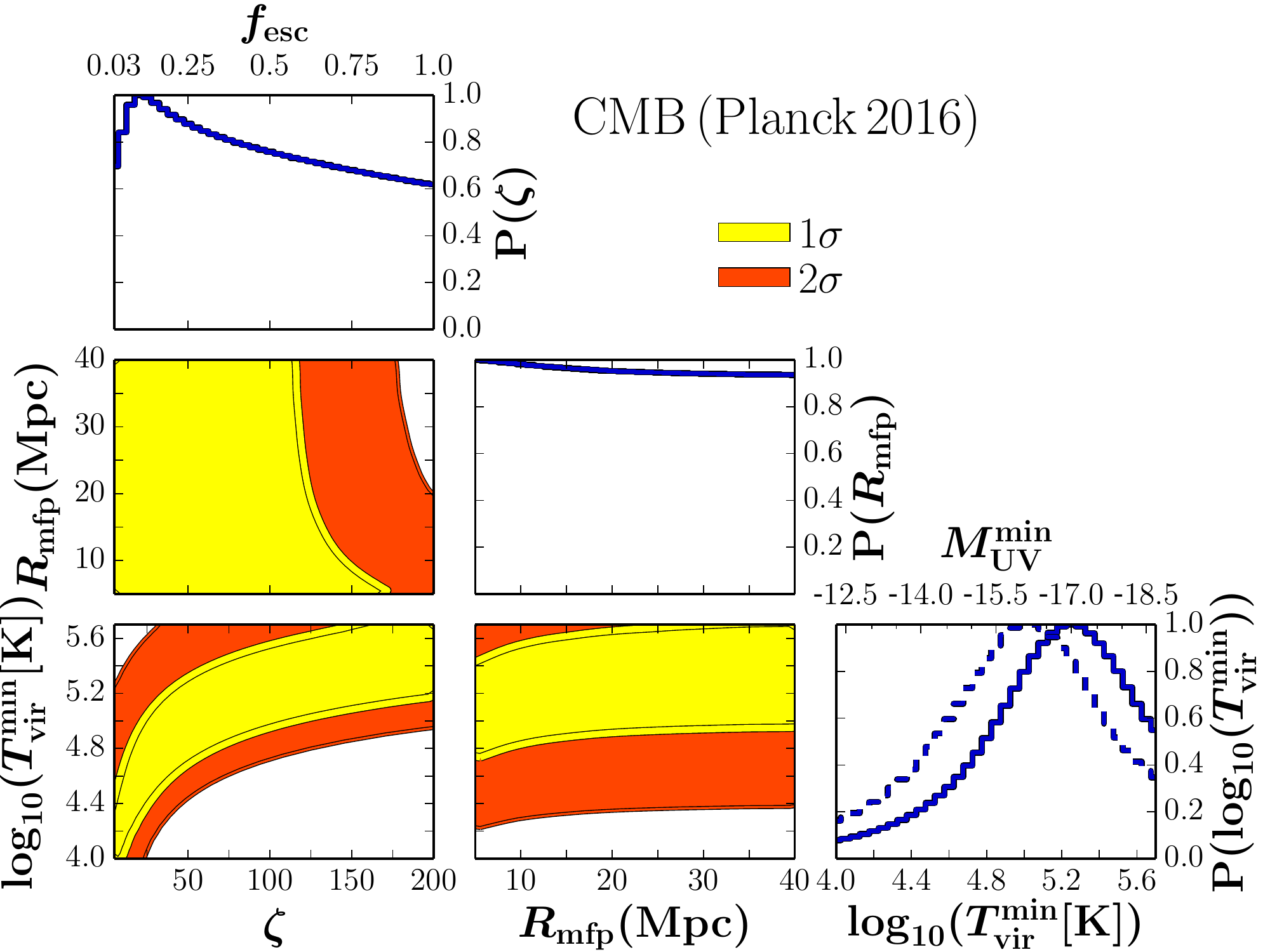}
}
\caption{
Same as Fig. \ref{fig:darkfraction}, but instead adopting the \citet{Planck16} model prior of $\tau_e = 0.058 \pm 0.012 ~ (1\sigma)$.
We caution that the seemingly well constrained value of the virial temperature is very sensitive to the adopted range of $\zeta$ priors, due to the fact that the likelihood is flat along a strip in the $\log(\Tvirmin)$--$\zeta$ plane; narrowing the allowed range for the ionising efficiency somewhat arbitrarily to $0<\zeta<100$ shifts the 1D marginalised $\Tvirmin$ PDFs to smaller values (c.f. the dashed, blue curve in the bottom right panel).
}
\label{fig:planck}
\vspace{-0.5\baselineskip}
\end{figure*}

The second observation used in our ``Gold Sample'' of model-independent priors comes from the Compton scattering optical depth to the CMB.  Photons reaching us from the Last Scattering Surface (LSS) get scattered along the way by free electrons.  This damps the primary temperature anisotropies, and introduces a large-scale polarisation signal.  The strength of this effect is quantified by the average Compton scattering optical depth, $\tau_e$, which is an integral measure of reionisation, integrating over the electron column density to the LSS.

The {\it Planck} satellite recently published updated constraints on $\tau_e$ from the 2016 data release \citep{Planck16}.  They provide marginalised limits from various data sets (Eqs. 4--7)\footnote{The choice of data sets and likelihood estimator, as well as the EoR model results in a $\sim10$ per cent scatter on the best-fit value of $\tau_e$ and its error bars \citep{Planck16}.   Moreover, the bias of the cosmic \hii{} regions during different patchy EoR models also causes a $\sim4$~per cent variation in $\tau_e$ \citep{MFS13}.
  Nevertheless, we somewhat generously include $\tau_e$ in the Gold Sample, as it is only weakly model-dependent, compared with the estimates below.}.
  The flagship constraint they quote comes from the high frequency instrument polarisation and Planck temperature data: $\tau_e = 0.058 \pm 0.012 ~ (1\sigma)$.

  In Fig. \ref{fig:planck} we show the impact of this {\it Planck} prior on the reionisation history\footnote{We note that the average midpoint of reionisation is lower than the instantaneous reionisation redshift, since EoR histories driven by the growth of dark matter structure are asymmetric, with a tail extending towards high-$z$.} and our EoR model parameters.  Contrary to the dark fraction prior, the Plank measurement discriminates most strongly against early reionisation models. More precisely, the integral constraint of $\tau_e$ limits the allowed parameter space to lie on roughly a diagonal strip of equal probability ($\chi^2 \approx 0$) in the $\Tvirmin$--$\zeta$ plane.  Marginalising the likelihood over this diagonal strip causes the 1D $\Tvirmin$ PDF to appear peaked. We caution that this is merely driven by the chosen range of priors; for example, if we narrow the allowed range of the ionising efficiency to more reasonable values, $0<\zeta<100$, the 1D $\Tvirmin$ PDF shifts to smaller values (compare the dashed and solid lines in the right-most panel of Fig. \ref{fig:planck}). The strong degeneracy between $\Tvirmin$ and $\zeta$ prevents robust constraints on astrophysical parameters from the $\tau_e$ measurement.

\subsection{Combined Gold sample}

\begin{figure*}
{
\includegraphics[trim = 1.1cm 0.7cm 0cm 0.3cm, scale = 0.445]{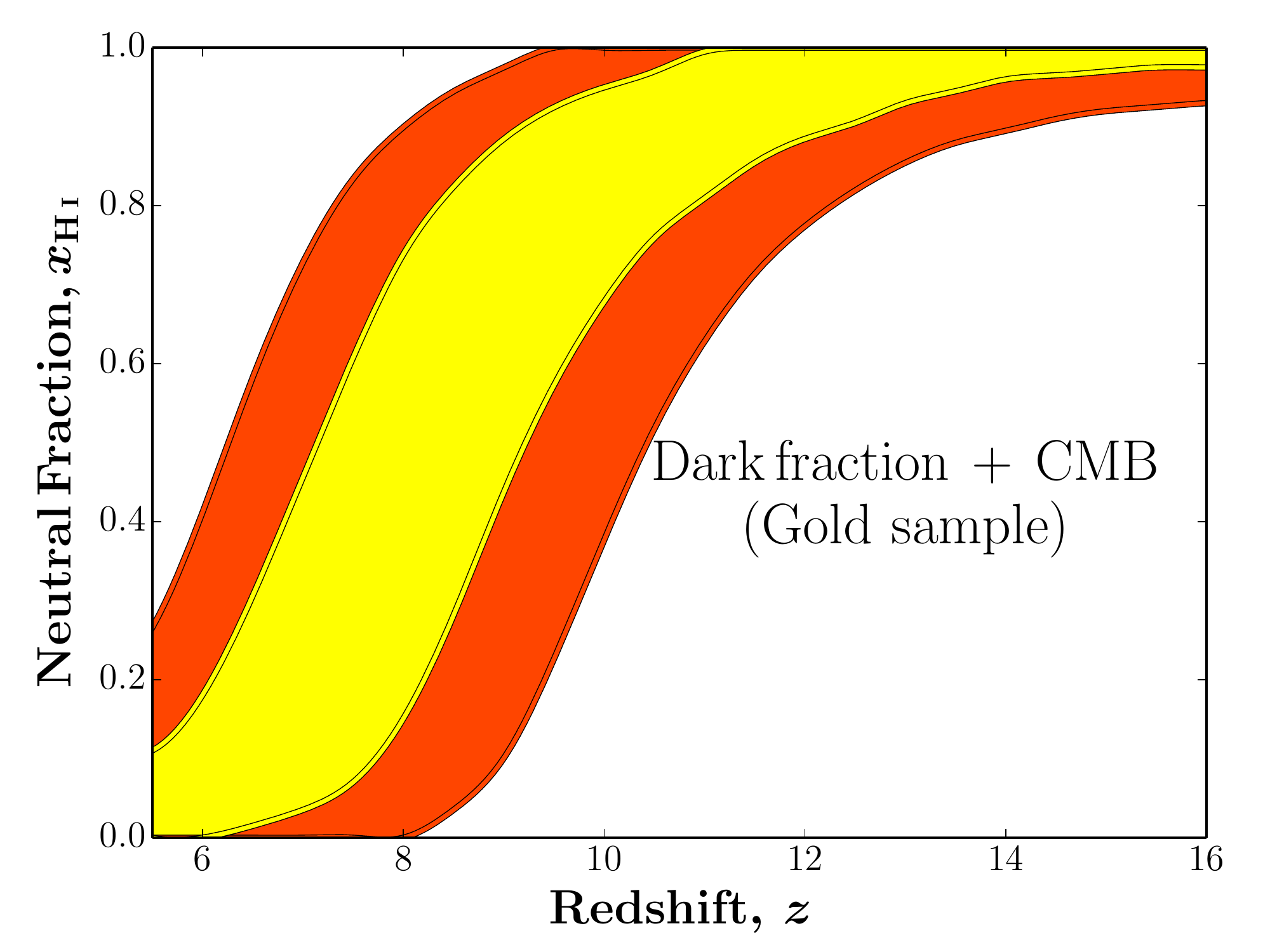}
\includegraphics[trim = 0.8cm 0.2cm 0cm 0.3cm, scale = 0.455]{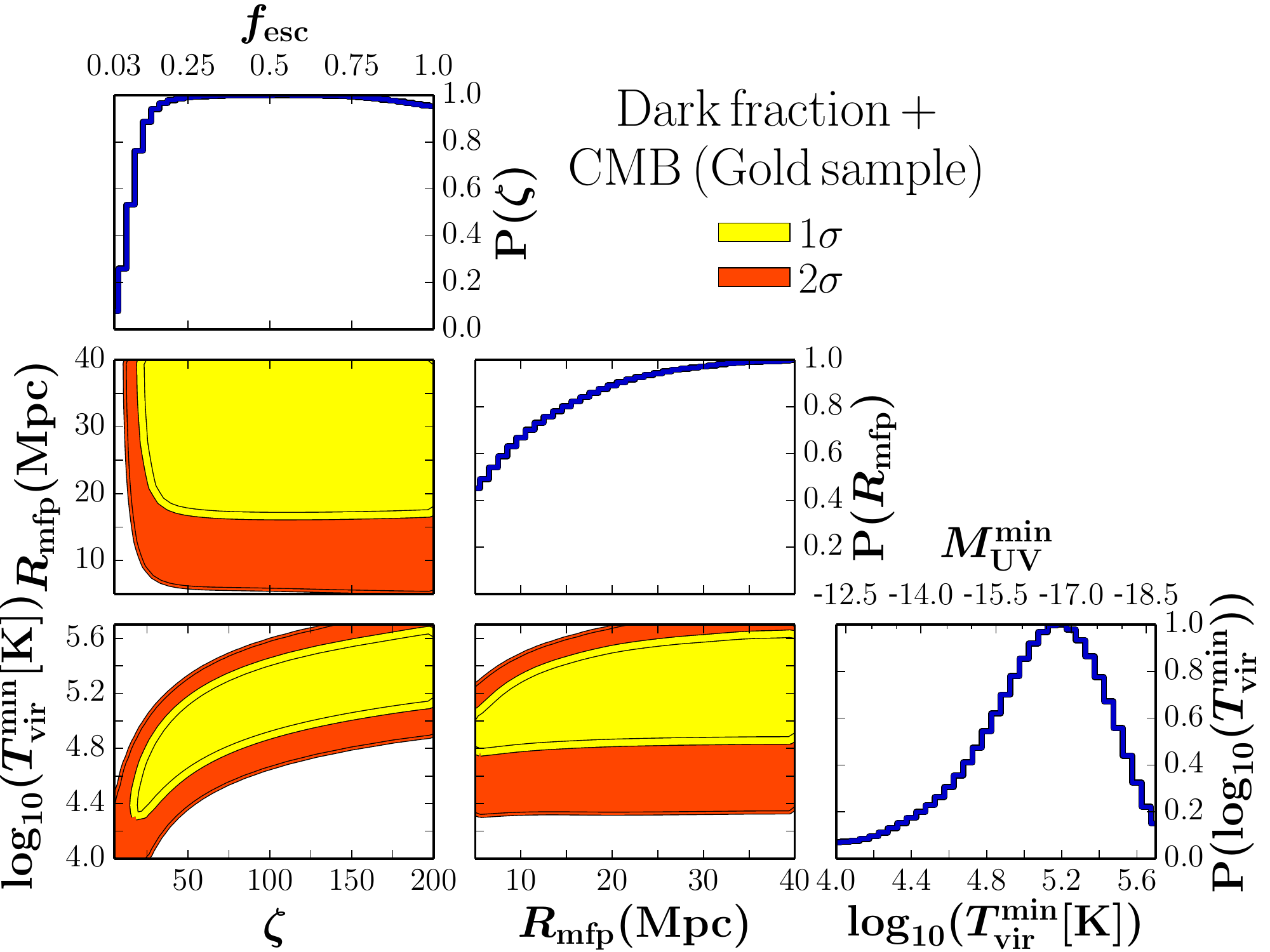}
}
\caption{
Same as Fig. \ref{fig:darkfraction}, but including both of the Gold Sample priors: (i) the dark fraction constraint of $\avenf(z=5.9) \lsim 0.06 + 0.05~(1\sigma)$; and (ii) the {\it Planck} 2016 constraint of $\tau_e = 0.058 \pm 0.012 ~ (1\sigma)$.
}
\label{fig:gold}
\vspace{-0.5\baselineskip}
\end{figure*}

In Fig. \ref{fig:gold} we show the combined impact of our two ``Gold Sample'' priors: (i) the dark fraction constraint of $\avenf(z=5.9) \lsim 0.06 + 0.05~(1\sigma)$; and (ii) the {\it Planck} 2016 estimate of $\tau_e = 0.058 \pm 0.012 ~ (1\sigma)$.  It is interesting to note that the reionisation history is already constrained by these complimentary priors.  As we shall see below, the addition of more uncertain, model-dependent constraints only moderately tightens these constraints.

The Gold Sample constraints on our model parameters are also interesting, though not robust.  Very slow reionisation histories (low $\zeta$, low $\mfp$) are disfavoured, as well as asymmetric reionisation histories with very extended end stages (low $\mfp$).  These models run into difficulty reionising the Universe late enough to match $\tau_e$, but also ending sufficiently early to match the dark fraction upper limit at $z=5.9$.  As discussed above, the constraint on $\Tvirmin$ is driven by the integral Planck constraints, and is sensitive to the adopted prior over $\zeta$.

Note that while our simple three parameter EoR model suffices to provide a physically-intuitive basis set of reionisation histories,
interpreting constraints on the astrophysical parameters themselves is less straightforward.  The EoR parameters should only be treated as ``effective'' parameters, since they average over redshift and halo mass evolution.

\section{Model-dependent EoR observational priors}
\label{sec:model_depend}

We now fold-in observational constraints which are more dependent on the EoR modelling.  We select the latest, most conservative (i.e. secure) results from studies tailored to each observable. We do however admit a bias in the selection of the data sets below.  For example, we do not include relatively-popular constraints based on the afterglow spectra of Gamma-Ray bursts (e.g. \citealt{Totani06, Chornock13}), as these do not (yet) have statistical significance in constraining {\it inhomogeneous} reionisation (e.g. \citealt{McQuinn08, MF08damp}).

\subsection{Lyman alpha emission from galaxies}

\subsubsection{The Lyman alpha fraction}

\begin{figure*}
{
\includegraphics[trim = 1.1cm 0.5cm 0cm 0.3cm, scale = 0.445]{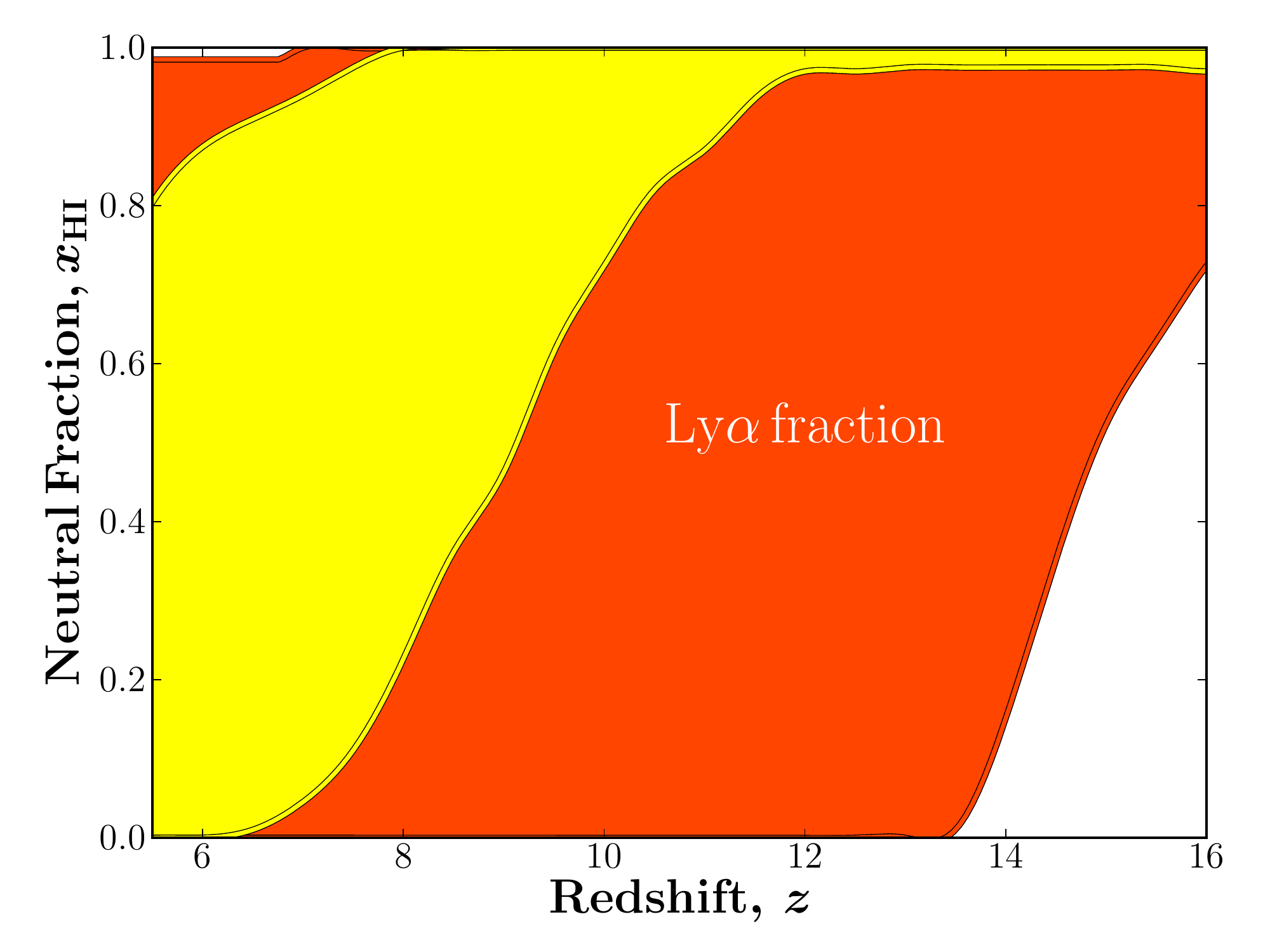}
\includegraphics[trim = 0.8cm 0cm 0cm 0.3cm, scale = 0.455]{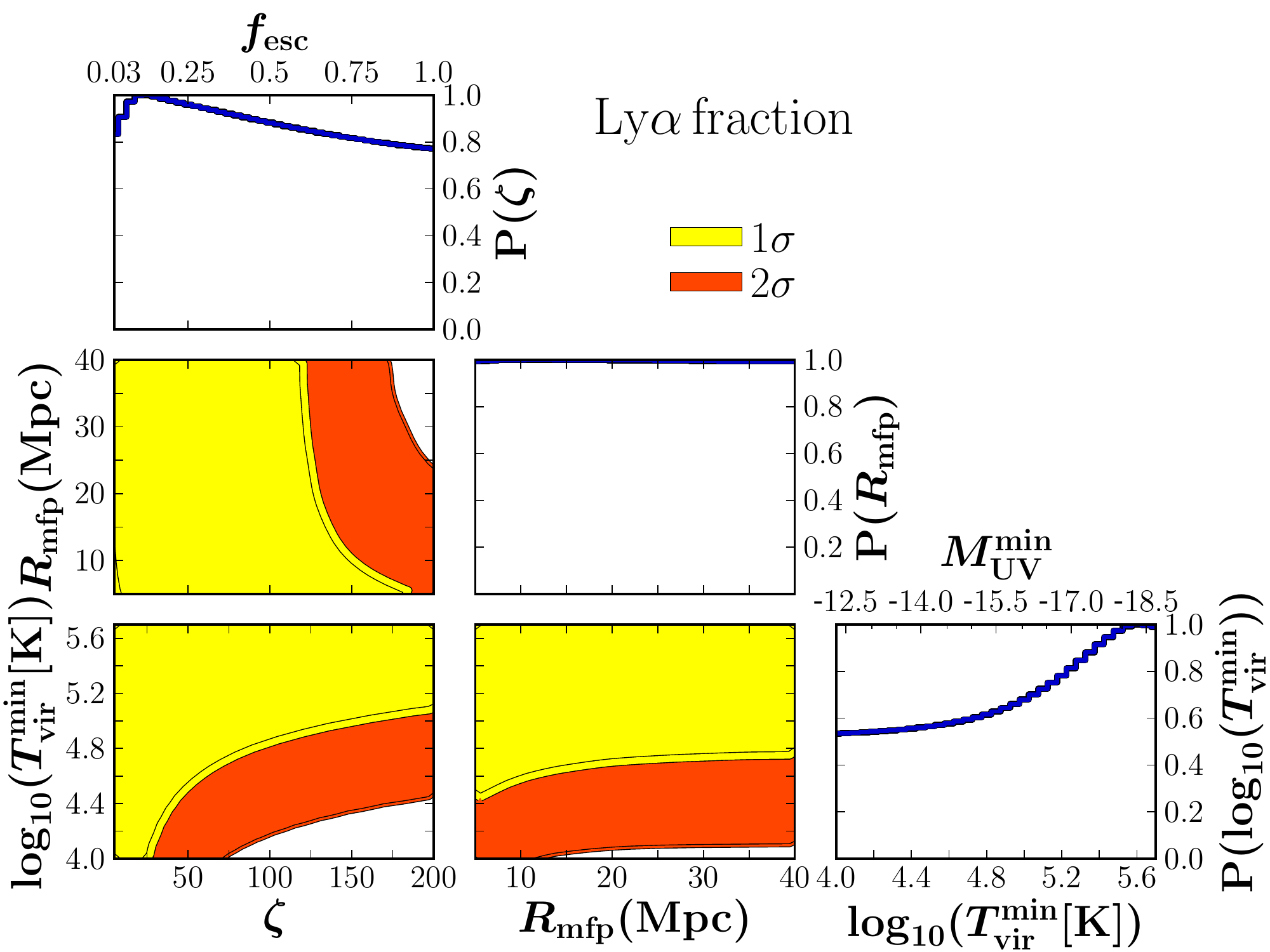}
\includegraphics[trim = 1.1cm 0.7cm 0cm 0cm, scale = 0.445]{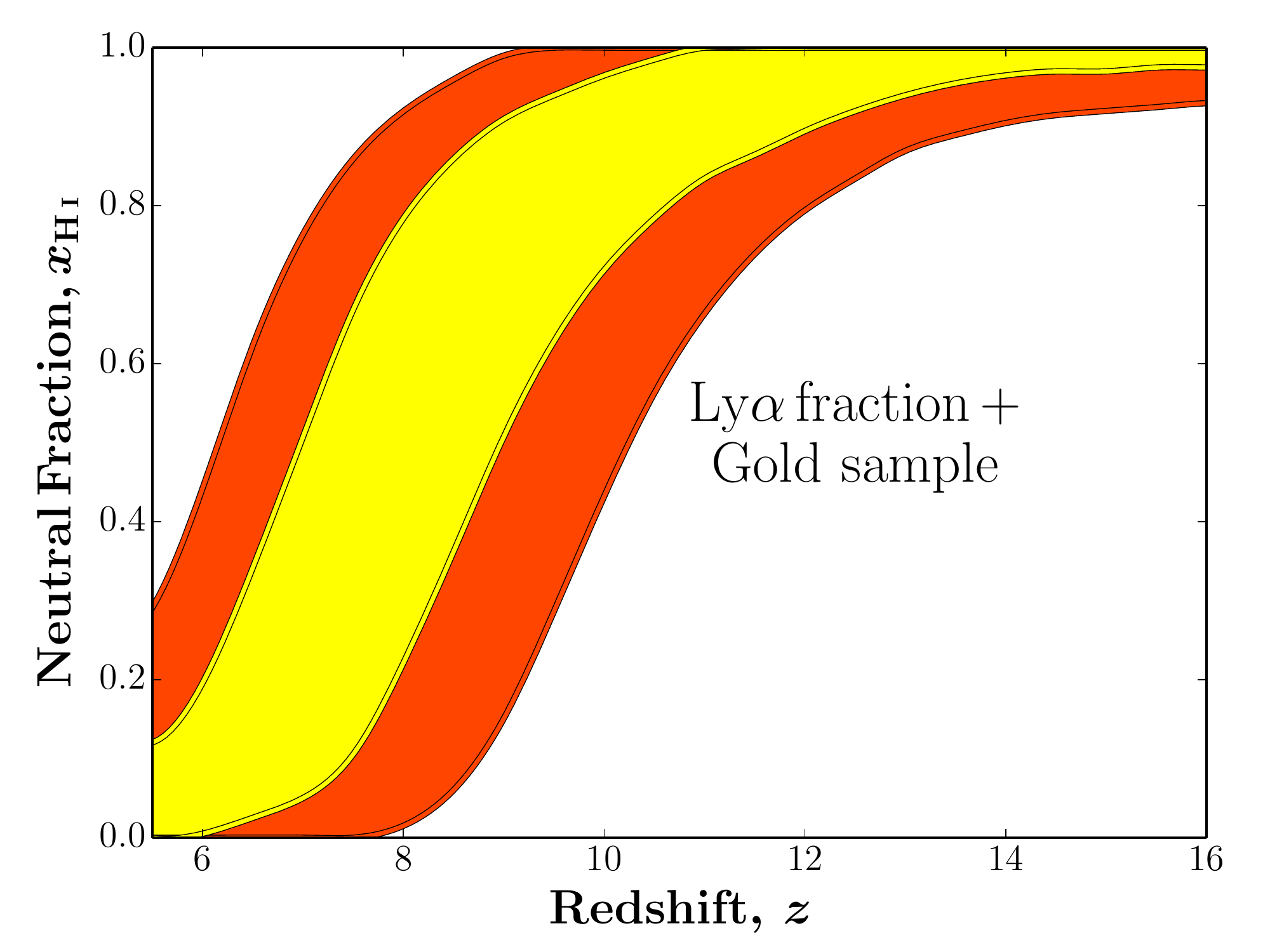}
\includegraphics[trim = 0.8cm 0.2cm 0cm 0cm, scale = 0.455]{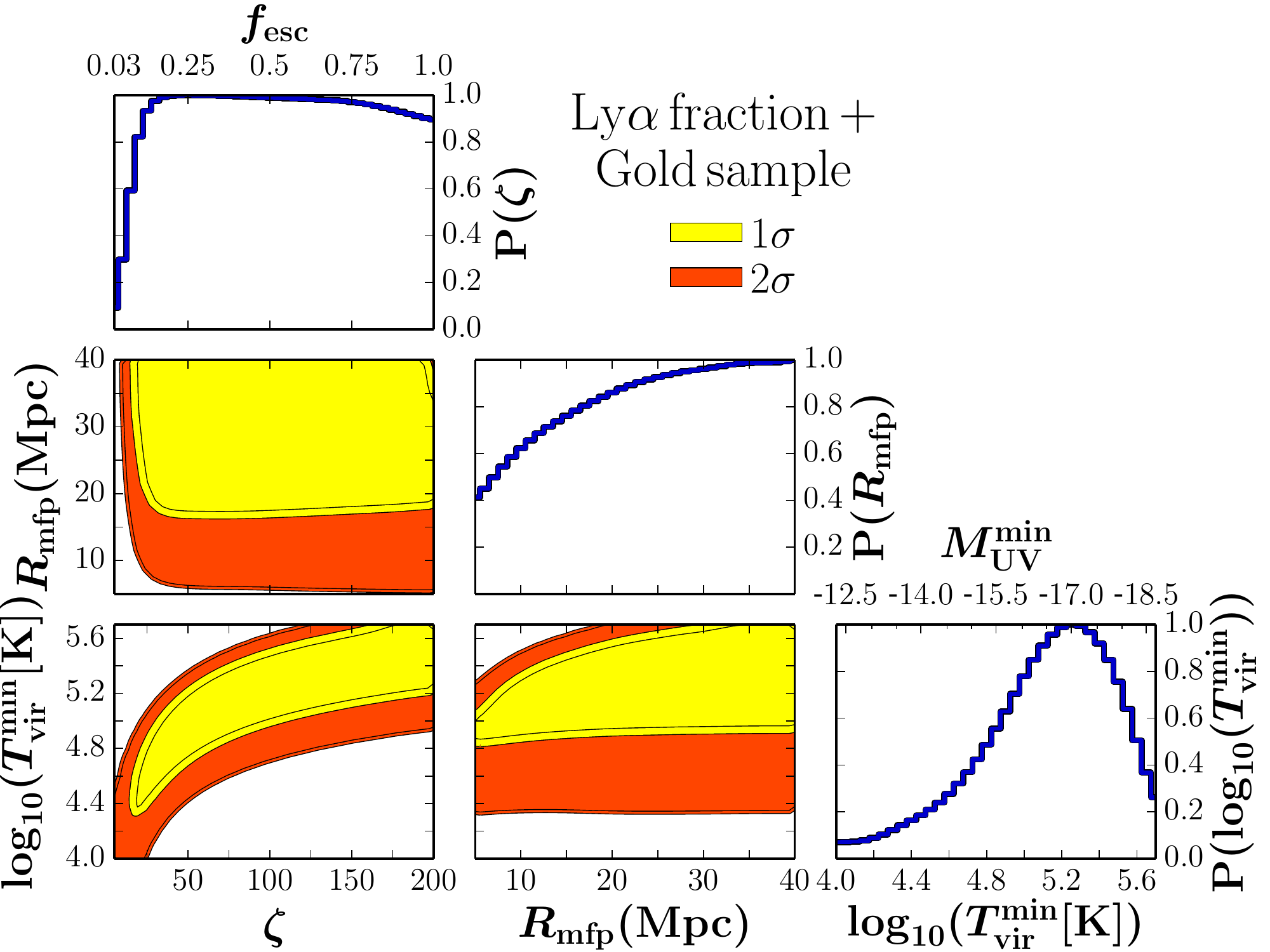}
}
\caption{
{\it Top row:} Same as Fig. \ref{fig:darkfraction}, but instead adopting the \lya\ fraction motivated prior of $\avenf(z=7)-\avenf(z=6) \geq 0.4$ from \citet{Mesinger15}.
{\it Bottom row:} Same as the top row, but additionally including the Gold Sample priors.
}
\label{fig:LAf}
\vspace{-0.5\baselineskip}
\end{figure*}

The \lya\ emission line emerging from galaxies during the EoR could be strongly attenuated by the cosmic \hi{} in the intervening IGM, primarily due to absorption in the damping wing of the \lya\ profile. Recently, several groups have noted that the fraction of colour-selected galaxies with a detectable \lya\ emission line (the so-called \lya\ fraction) seems to drop dramatically beyond $z>6$ (e.g. \citealt{Stark10, Pentericci11, Ono12, Caruana14, Schenker14}).  Although still limited by small number statistics, these results are very suggestive of a strong evolution in the neutral hydrogen fraction over the redshift interval $z\approx$~6--7 (e.g. \citealt{McQuinn07LAE, MF08LAE, DMW11, Jensen13, Dijkstra14}).

Quantitatively interpreting the observations is difficult, as it requires accounting for the uncertainties in the intrinsic galactic emission profile, the ionising background inside local \hii{} regions (which determines the incidence of high-column density systems), and the large-scale morphology of the EoR.  For this work, we take the statistical constraints from \citet{Mesinger15}, who used a tiered model with three components: (i) analytic intrinsic emission profiles; (ii) hydrodynamic simulations of the self-shielded systems in the local IGM; (iii) semi-numerical simulations of the large-scale reionisation field.  Adopting conservative priors and marginalising over uncertainties in the ionising background, these authors obtained a 1$\sigma$ upper limit of $\avenf(z=7)\geq0.4$\footnote{This result is roughly consistent with the subsequent, similar analysis of \citet{Choudhury15}; however, here we only take the results from the former work as the latter does not provide statistically-quantitative constraints nor does it account for uncertainties in the ionising background.}.  
It is important to note that the $z\approx7$ neutral fraction constraint in \citet{Mesinger15}, {\it assumed} that the Universe was reionised by $z=6$, since it is the {\it relative change} of $\avenf$ over $z=6\rightarrow7$, which drives the observed drop in the \lya\ fraction.  To relax this assumption, here we instead require that the neutral fraction evolves by $\gsim$ 40 per cent over the redshift interval $z=6\rightarrow7$, i.e.  $\Delta_{7-6} \avenf \equiv \avenf(z=7)-\avenf(z=6) \geq 0.4$. We implement this constraint by adopting a one-sided Gaussian prior for $\Delta_{7-6} \avenf$, with a peak at $\Delta_{7-6} \avenf=1$ and a $\sigma$ of $\Delta_{7-6} \avenf =0.6$.

We note that the change in the neutral fraction required to match the \lya\ fraction observations should decrease as $\avenf(z=6)\rightarrow1$; however, this area of parameter space is already disfavoured by the dark fraction observations and so does not quantitatively impact our final conclusions.

In Fig. \ref{fig:LAf}, we show the constraints on the reionisation history and EoR model parameters, provided by the \lya\ fraction motivated prior, $\Delta_{7-6} \avenf \gsim 0.4$ ({\it top row}).  The models which are most consistent with such a rapid, late EoR evolution lie in the high-$\Tvirmin$ region of parameter space, as this parameter most strongly influences the timing of reionisation.  Adding the Gold Sample priors in the bottom row, we see that the resulting constraints are quite similar to those from the Gold Sample itself, albeit with a notable decrease in the peak model likelihood: from $\chi^2 =0$ in the Gold Sample, to $\chi^2=1.6$ when adding in the \lya\ fraction prior (see Table 1).  Indeed this \lya\ fraction constraint is mildly in tension with other observations, driving down the likelihood of the best fit model using all constraints combined (as we shall see below).  Additionally including evolution in galaxy properties might loosen the required evolution in $\avenf$ (e.g. \citealt{Finkelstein12, Dijkstra14, Mesinger15, Choudhury15}); future studies will focus on doing this in a non ad-hoc manner.

\subsubsection{Clustering of Lyman alpha emitters} \label{sec:LAEclust}

\begin{figure*}
{
\includegraphics[trim = 1.1cm 0.5cm 0cm 0.3cm, scale = 0.445]{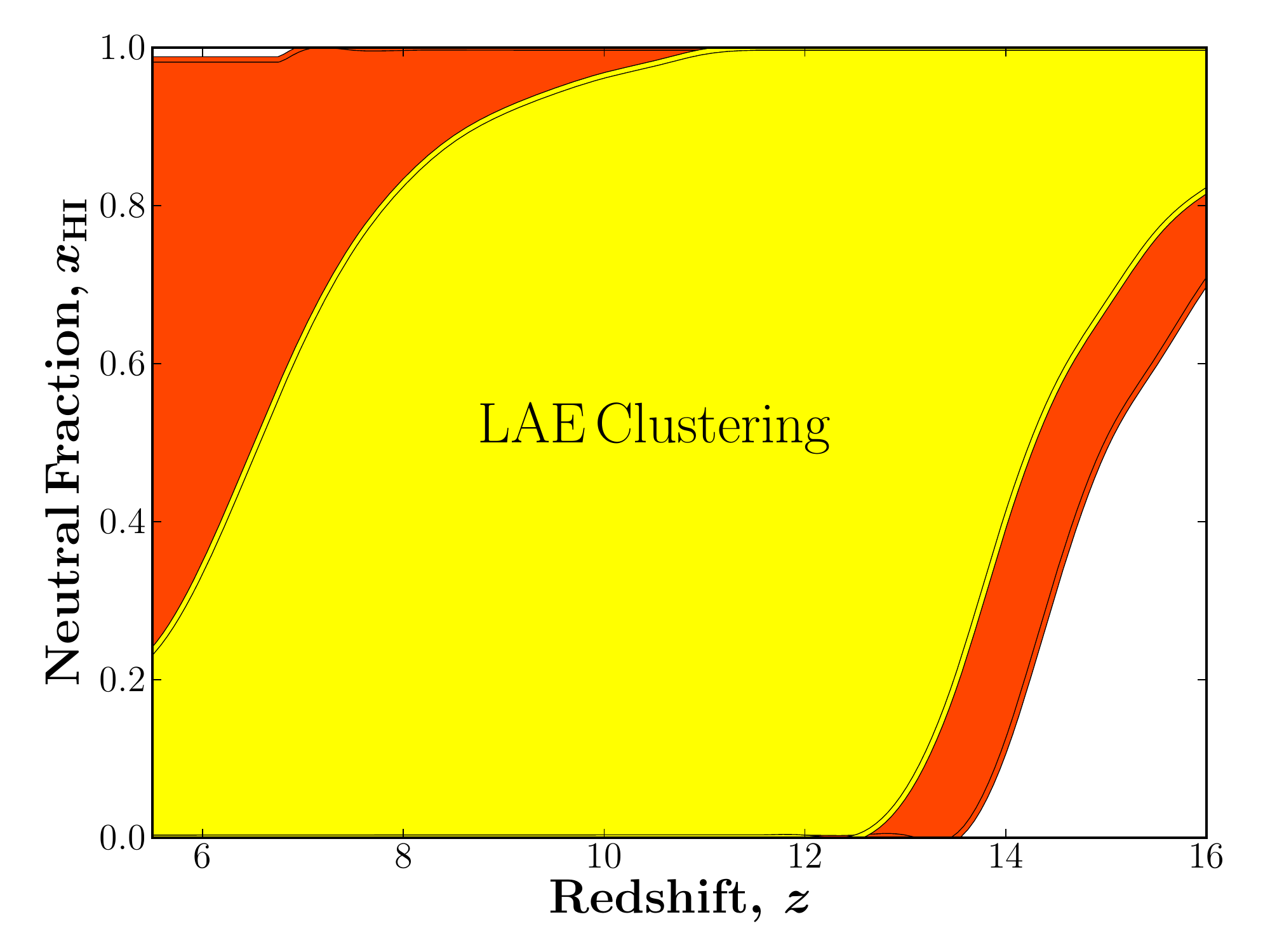}
\includegraphics[trim = 0.8cm 0cm 0cm 0.3cm, scale = 0.455]{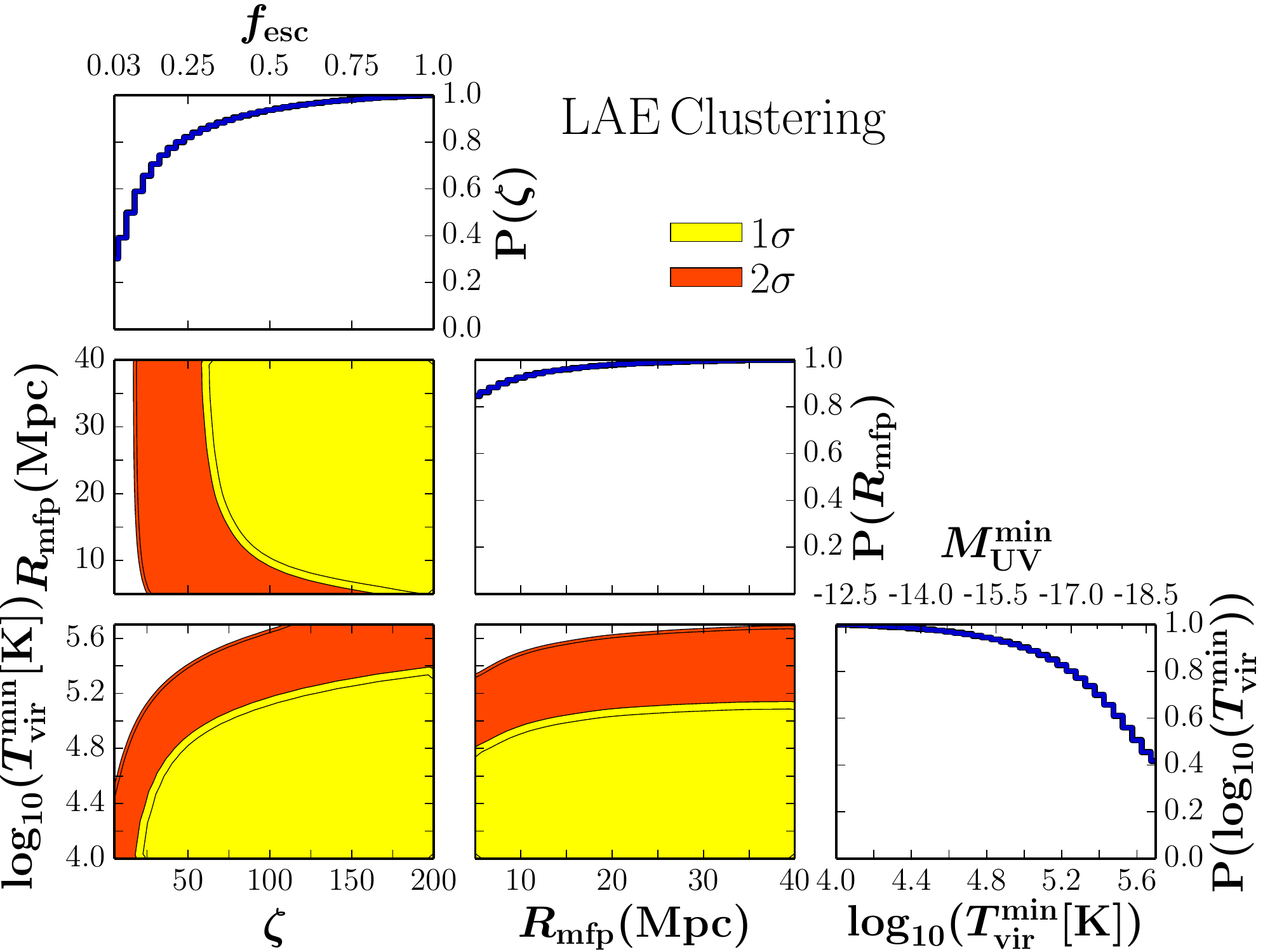}
\includegraphics[trim = 1.1cm 0.7cm 0cm 0cm, scale = 0.445]{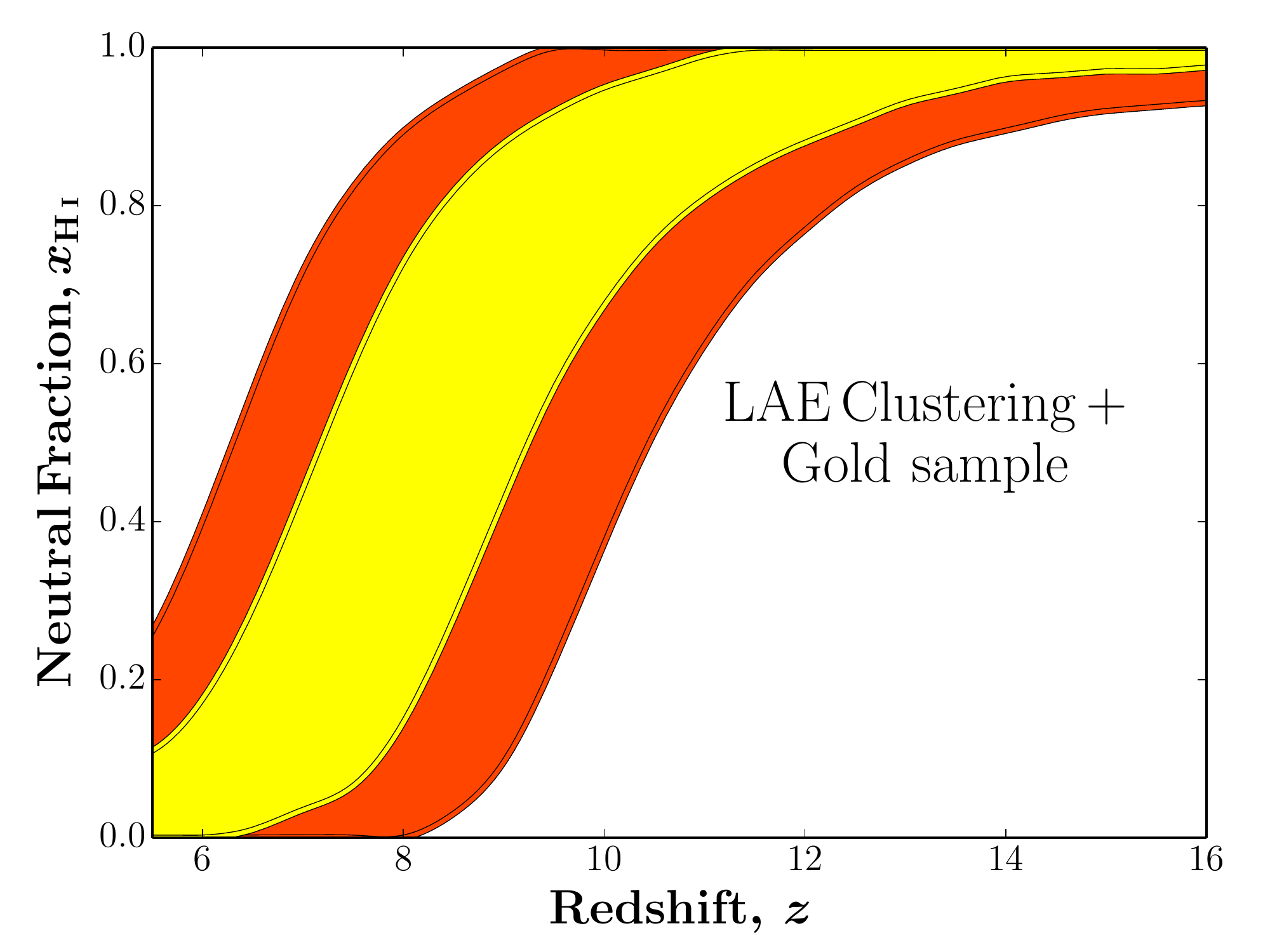}
\includegraphics[trim = 0.8cm 0.2cm 0cm 0cm, scale = 0.455]{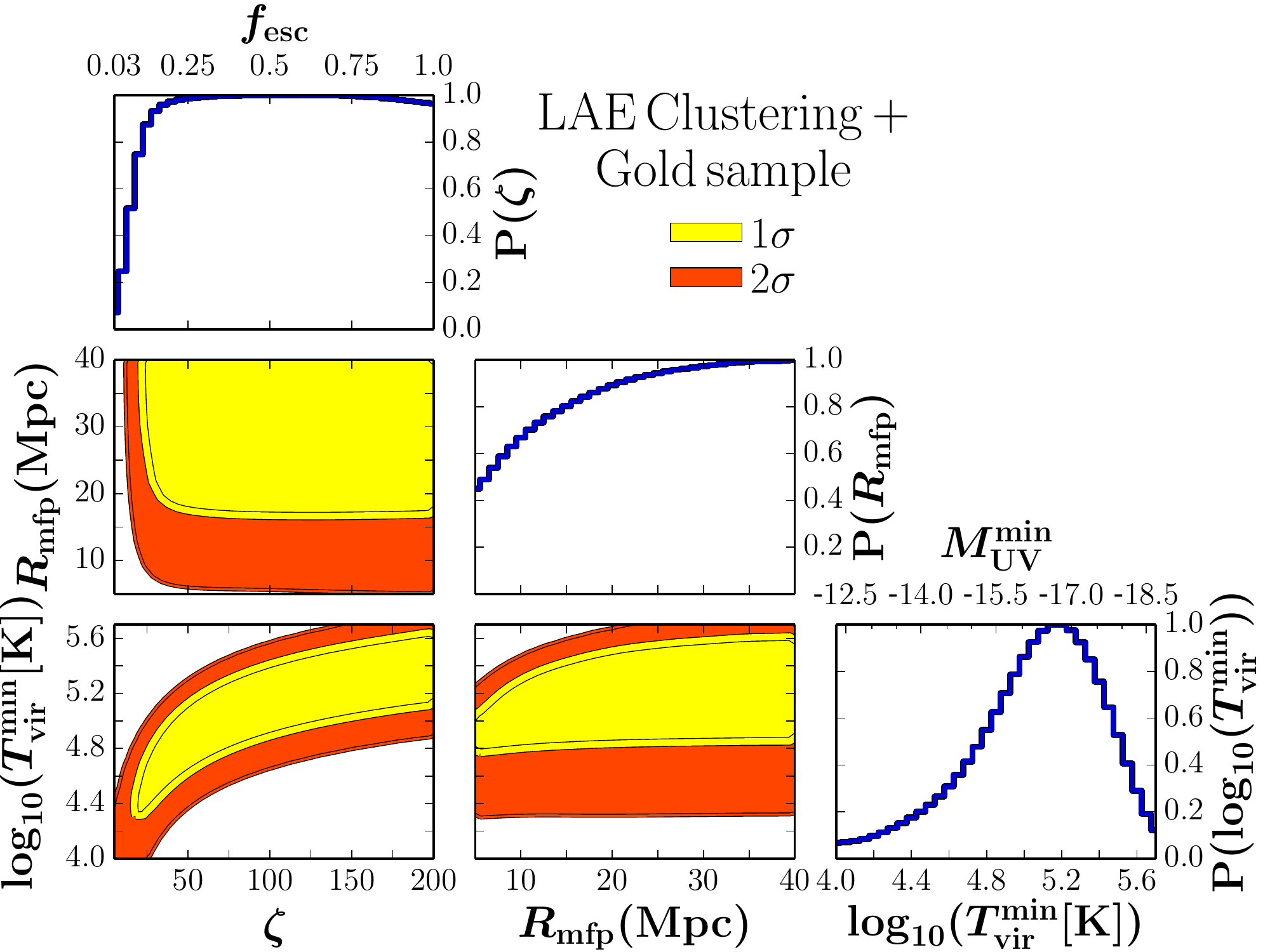}
}
\caption{
Same as Fig. \ref{fig:LAf}, but adopting instead the one-sided Gaussian prior of $\avenf(z=6.6) \leq 0.5 ~ (1\sigma)$, motivated by the observed LAE angular correlation function.
}
\label{fig:LAE_clustering}
\vspace{-0.5\baselineskip}
\end{figure*}

The strength of the \lya\ absorption from the IGM depends on the spatial distribution of galaxies and cosmic \hi{} patches.   Galaxies residing closer to the cosmic \hi{} patches will experience more absorption.  This impacts the observed clustering of LAEs (e.g. \citealt{FZH06, McQuinn07LAE, MF08LAE}).  {\it When normalised to a fixed observed LAE number density}, the reionisation constraint from LAE clustering is more robust than constraints from the number evolution, due to a weaker dependence on the unknown intrinsic \lya\ emission emerging from the galaxy.   The local galactic environment (e.g. accretion flows, outflows, self-shielded systems) have a weaker spatial signature than the EoR absorption on large-scales ($\gsim10$ Mpc).  This means that the clustering signal is (almost) uniquely determined by just: (i) the typical dark matter host haloes of LAEs; and (ii) $\avenf$\footnote{The observed angular correlation function of LAEs at fixed $\avenf$ should, in principle, also depend on the EoR morphology.  However, narrow-band surveys with $\Delta z\approx0.1$ (corresponding to $\sim40$ Mpc) average over this structure, making the resulting correlation function very insensitive to the EoR morphology at fixed $\avenf$ \citep{SM15}.}.

Here we use the recent constraint of $\avenf(z=6.6) \leq 0.5$ (1-$\sigma$) derived by comparing the LAE angular correlation function (ACF) observed by Subaru \citep{Ouchi10}, to a conservatively broad range of modelled LAE ACFs during the EoR (\citealt{SM15}; see also \citealt{McQuinn07LAE, Ouchi10, Jensen13}).  Specifically, we adopt a one-sided Gaussian prior on $\avenf(z=6.6)$, with a peak value at zero and $\sigma=0.5$.

We show the resulting constraints on the reionisation history and model parameters in Fig. \ref{fig:LAE_clustering}. As could be expected, the trends are qualitatively similar to the ones corresponding to the dark fraction in  Fig. \ref{fig:darkfraction}: a preference against late reionisation models.  However the LAE clustering constraints are considerably weaker.  Thus their inclusion does not improve upon the constraints already obtained with the two Gold Sample priors (bottom panels).

\subsection{Damping wing in QSO spectra: constraint from ULAS J1120+0641}
\label{sec:QSO_damp}

\begin{figure*}
{
\includegraphics[trim = 1.1cm 0.5cm 0cm 0.3cm, scale = 0.445]{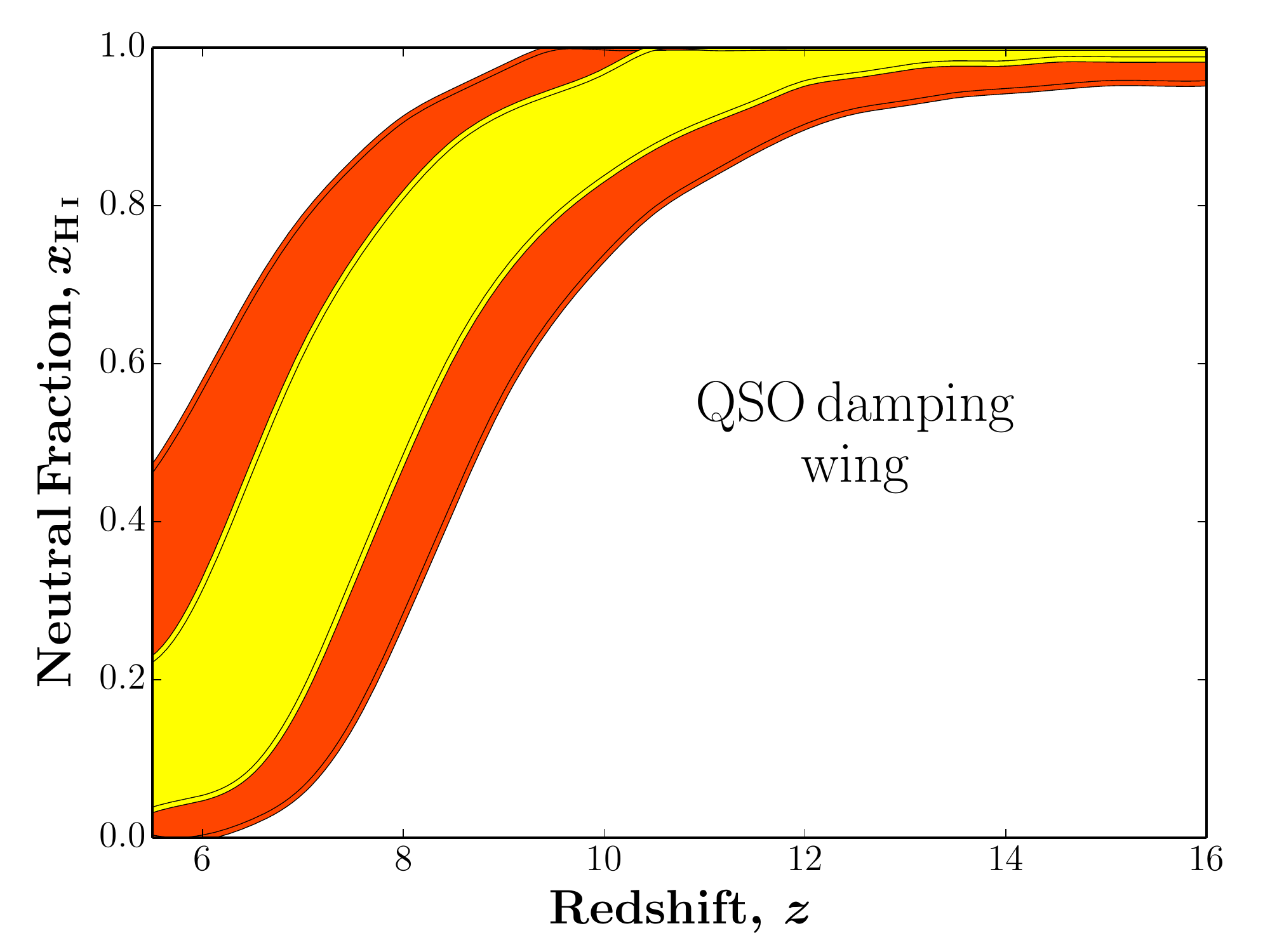}
\includegraphics[trim = 0.8cm 0cm 0cm 0.3cm, scale = 0.455]{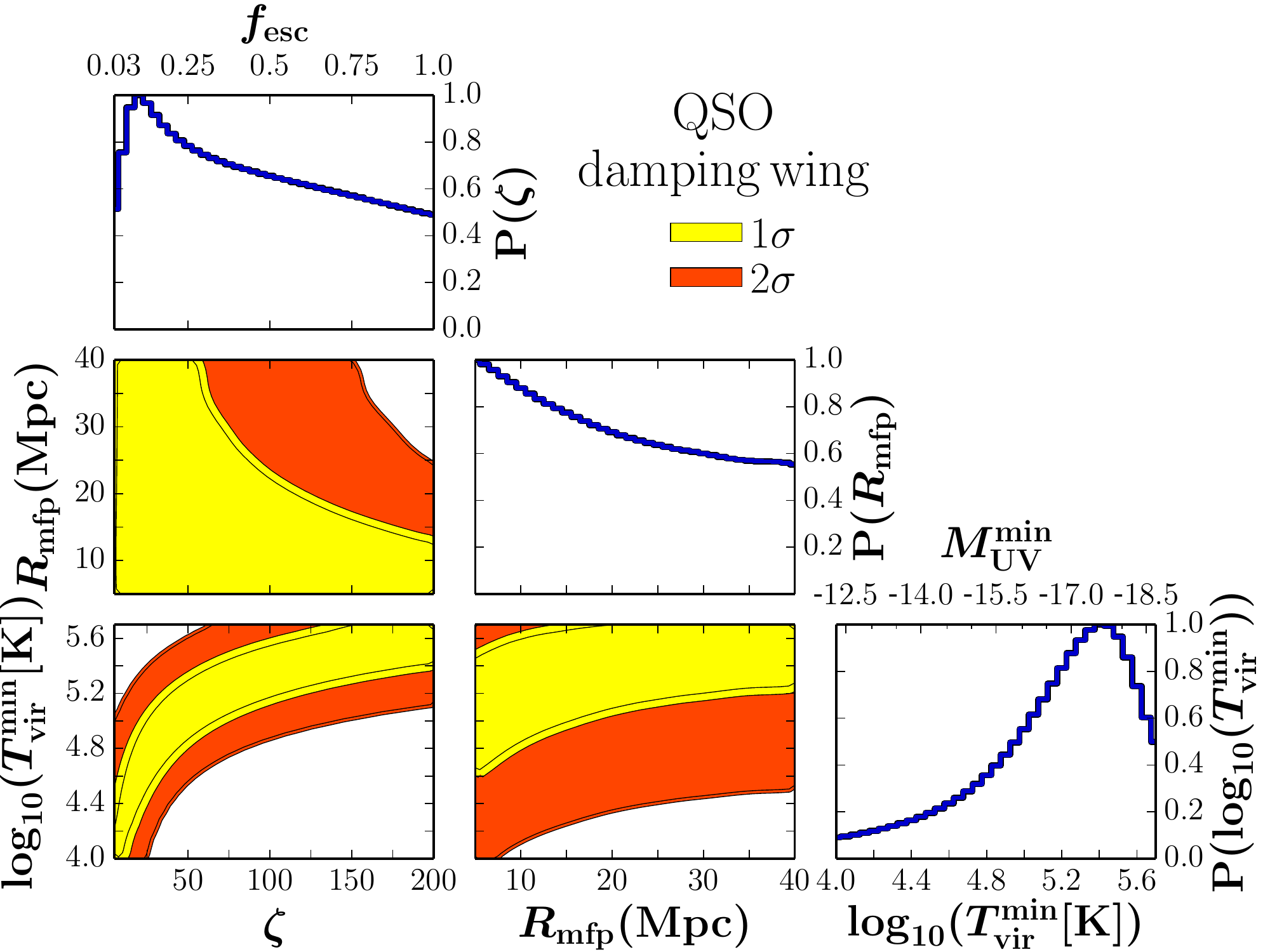}
\includegraphics[trim = 1.1cm 0.7cm 0cm 0cm, scale = 0.445]{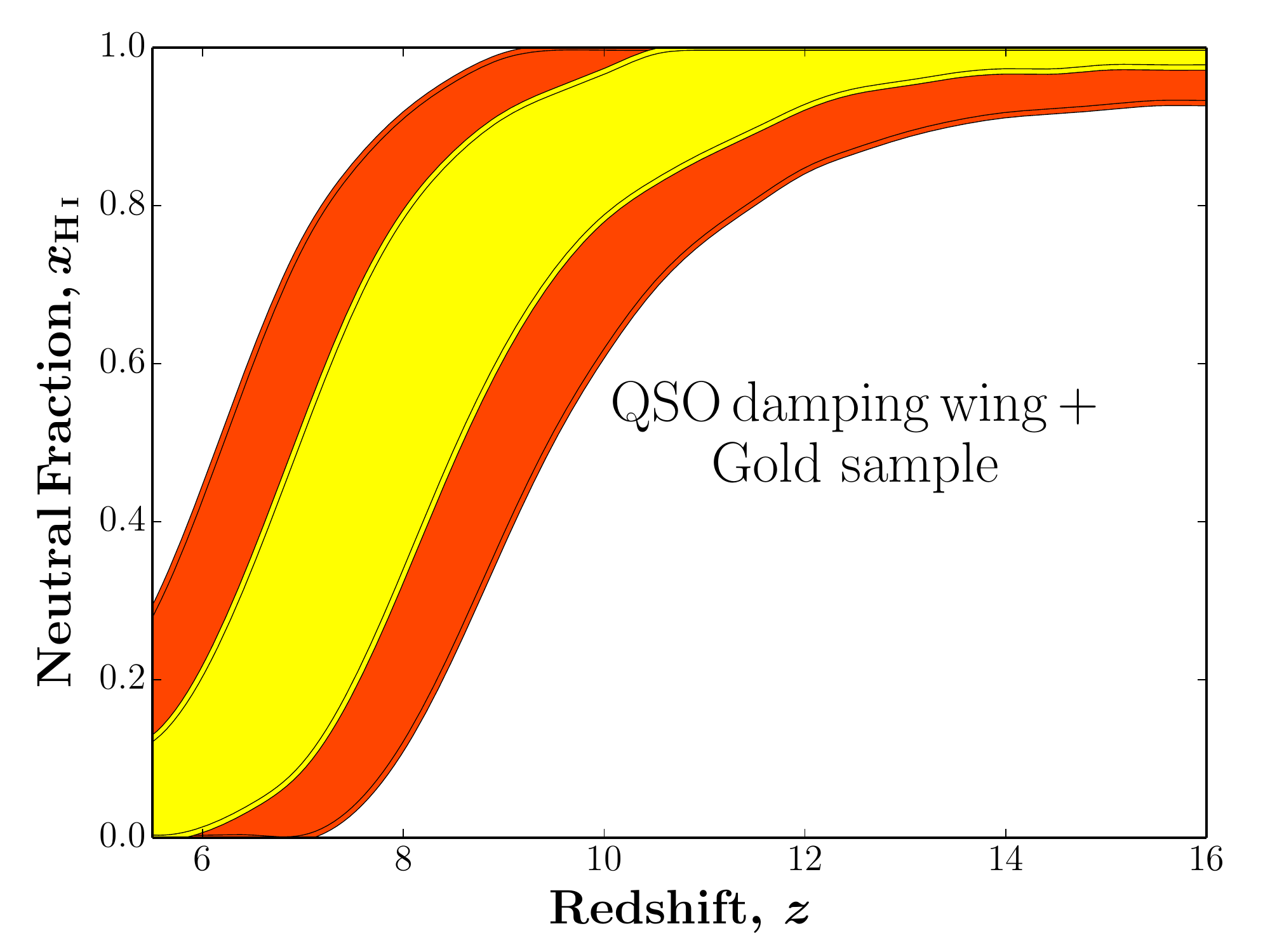}
\includegraphics[trim = 0.8cm 0.2cm 0cm 0cm, scale = 0.455]{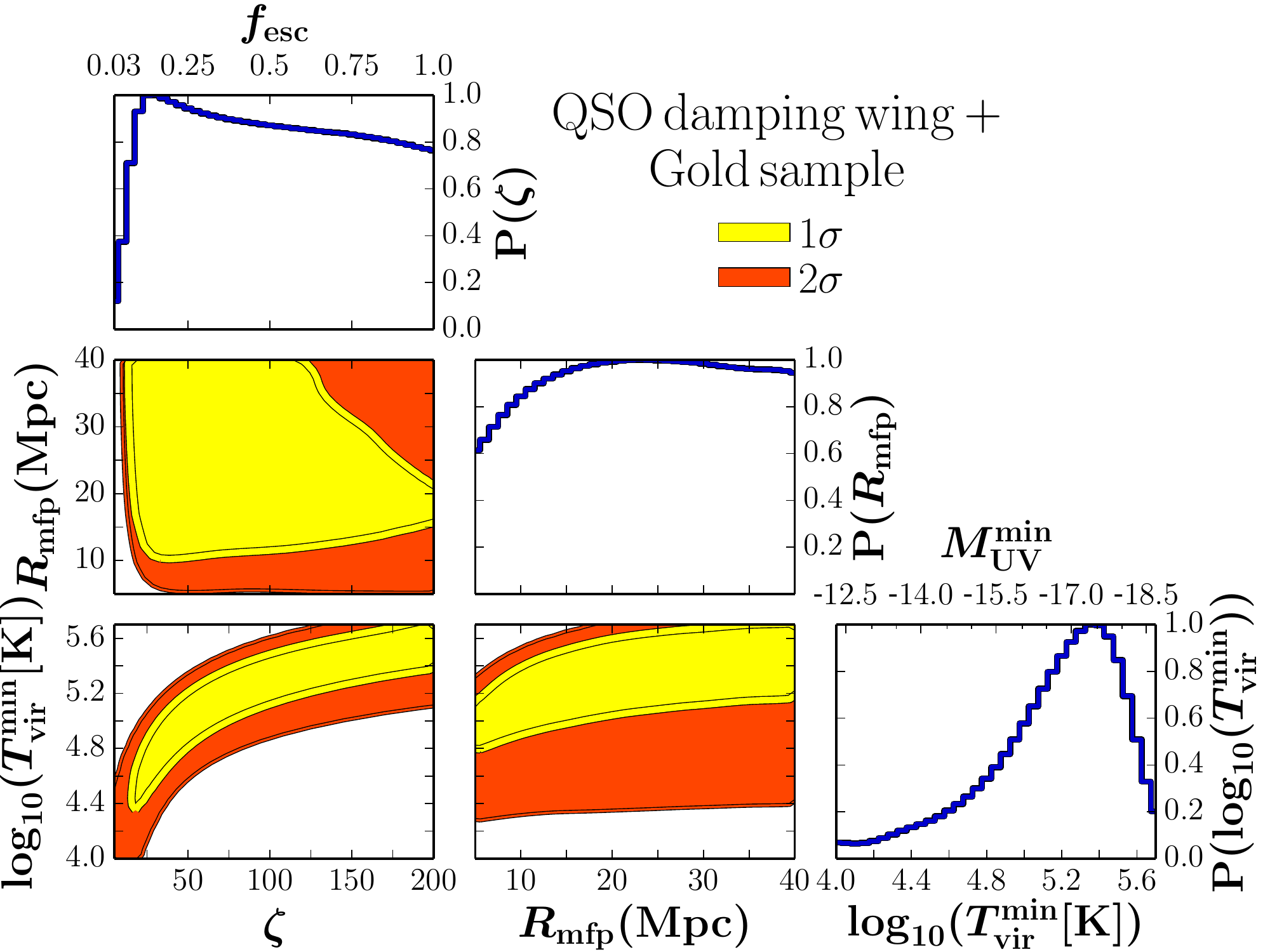}
}
\caption{
  Same as Fig. \ref{fig:LAf}, but adopting the $\avenf(z=7.1)$ prior from the analysis of ULAS J1120+0641 in \citet{Greig16}: $\avenf = 0.40^{+0.21}_{-0.19}$ at $1\sigma$ ($\avenf = 0.40^{+0.41}_{-0.32}$ at $2\sigma$).
}
\label{fig:QSO_damp}
\vspace{-0.5\baselineskip}
\end{figure*}

The imprint of damping wing absorption from cosmic \hi{} has also been studied in $z>6$ QSO spectra.  Studies quantifying this effect have focused on: (i) the distribution of flux in the so-called near zone: the spectral region blueward of the \lya\ line where the ionising flux from the QSO itself facilitates transmission \citep{MH04, MH07, SMH13}; and (ii) the amplitude of the observed \lya\ line \citep{Mortlock11, Bolton11}.  In the case of (i), one has to model the \lya\ (and/or \lyb) forests in the near zone, statistically searching for an additional smooth absorption component, corresponding to the damping wing, among the fluctuating forest (e.g. \citealt{MHC04}).
In the case of (ii), one does not have to model the forest, but must instead accurately quantify the degeneracy of the damping wing imprint and the uncertainties in the intrinsic (unabsorbed) emission profile (e.g. \citealt{BB15})\footnote{The intrinsic emission profile is less of a concern for the aforementioned studies of the flux distribution in the near zone, since those QSOs are surrounded by very large \hii{} regions.  Thus any damping wing absorption would be weak on the red side of the \lya\ line, allowing the red side to be used for estimating the unabsorbed flux blueward of the line \citep{KH09}.}.
In both cases, one must also quantify the degeneracy between the damping wing from cosmic \hi{} and that from a possible high-column density damped Lyman alpha system (DLA) inside the ionised IGM\footnote{In the case of ULAS J1120+0641, searches for metal lines corresponding to a potential DLA along the line of sight have been unsuccessful (e.g. \citealt{Simcoe12, Odorico13}).  Thus if the damping wing signature was indeed from a DLA, it would have to be uncharacteristically metal poor (e.g. \citealt{MCM13}).  In any case, such high-column density DLAs are extremely rare in randomly-chosen IGM patches (e.g. \citealt{POW10}), making them highly unlikely culprits for providing a damping wing signature.}.

Here we use the $\avenf$ likelihood 
presented in \citet{Greig16}. This study reconstructed the intrinsic \lya\ emission profile of the $z=7.1$ QSO ULAS J1120+0641 \citep{Mortlock11}, using a covariance matrix of high-ionisation emission line properties generated from $\sim2000$ BOSS spectra \citep{Greig16Meth}.  The \lya\ emission line was characterised by a double Gaussian, with each component described by a height, width and velocity offset. Following the pipeline outlined in \citet{Greig16Meth}, we recover a six dimensional likelihood function from which we sampled $10^5$ times to extract a distribution of reconstructed \lya\ profiles. Each of these reconstructed intrinsic emission profiles was then multiplied by $10^5$ \lya\ damping wing absorption profiles sampled from a large EoR simulation at different values of $\avenf$ \citep{MGS2016}.  These combined intrinsic emission and mock IGM absorption spectra were then compared against the observed FIRE spectrum of ULAS J1120+0641 \citep{Simcoe12}.  In \citet{Greig16} we found the observed spectrum favoured a partially neutral IGM: $\avenf = 0.40\substack{+0.21 \\ -0.19}$ at $1\sigma$ ($\avenf = 0.40\substack{+0.41 \\ -0.32}$ at $2\sigma$). These constraints were found to be insensitive to the assumed EoR topology, at a fixed value of $\avenf$.

The resulting constraints on the EoR history and astrophysical parameters are shown in Fig. \ref{fig:QSO_damp}.
{\it These are the strongest constraints currently available}, illustrating the usefulness of bright high-$z$ QSOs.  As was the case for the \lya\ fraction, late and rapid EoR evolution is preferred.  However, in contrast to the \lya\ fraction, these QSO damping wing constraints are not at odds with the Gold Sample: the best fit model for the combined priors (shown in the bottom panels of Fig. \ref{fig:QSO_damp}) has a very reasonable $\chi^2 = 0.23$.

\subsection{The patchy kinetic Sunyaev-Zel'dovich signal}

\begin{figure*}
{
\includegraphics[trim = 1.1cm 0.5cm 0cm 0.3cm, scale = 0.445]{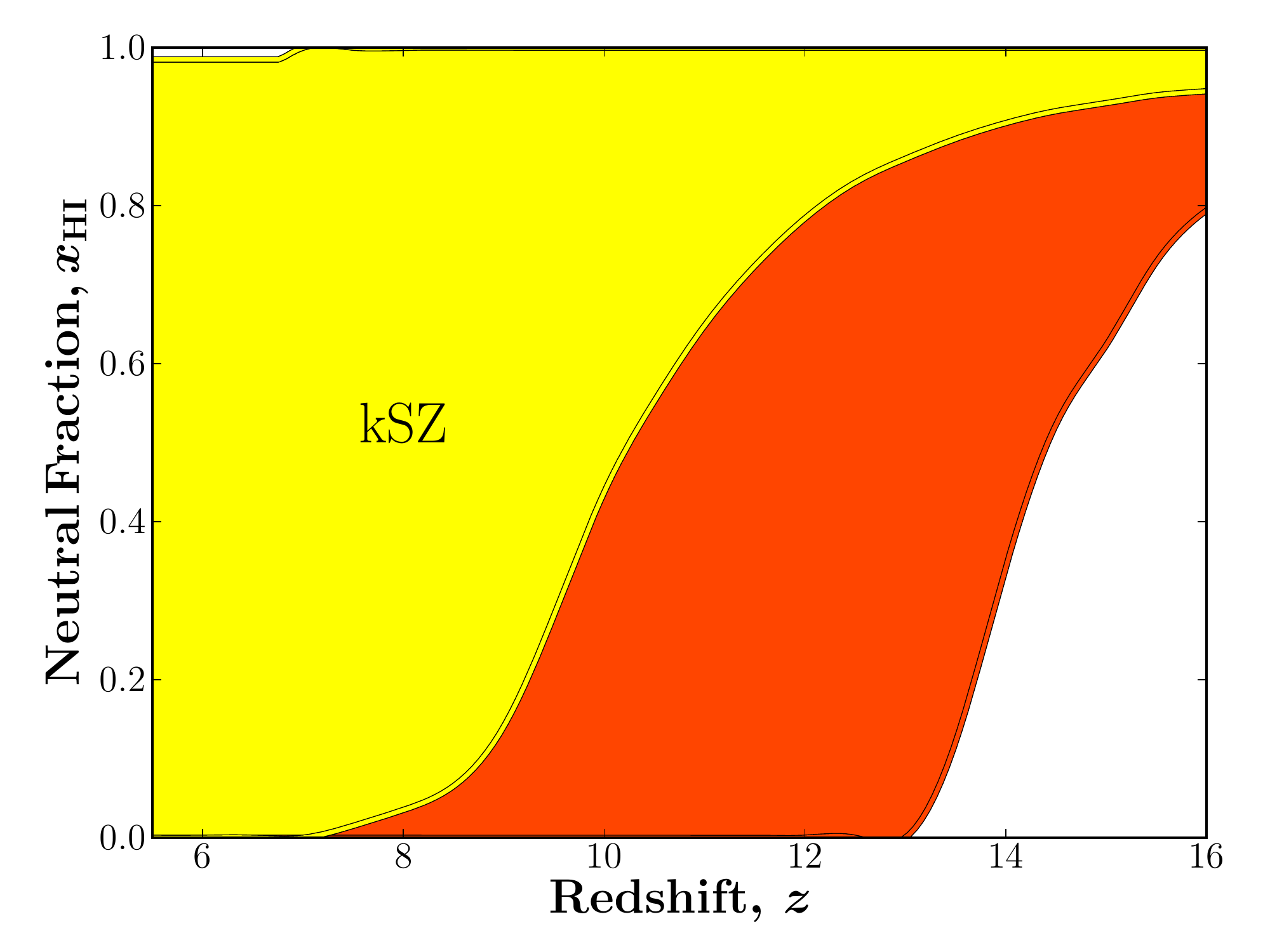}
\includegraphics[trim = 0.8cm 0cm 0cm 0.3cm, scale = 0.455]{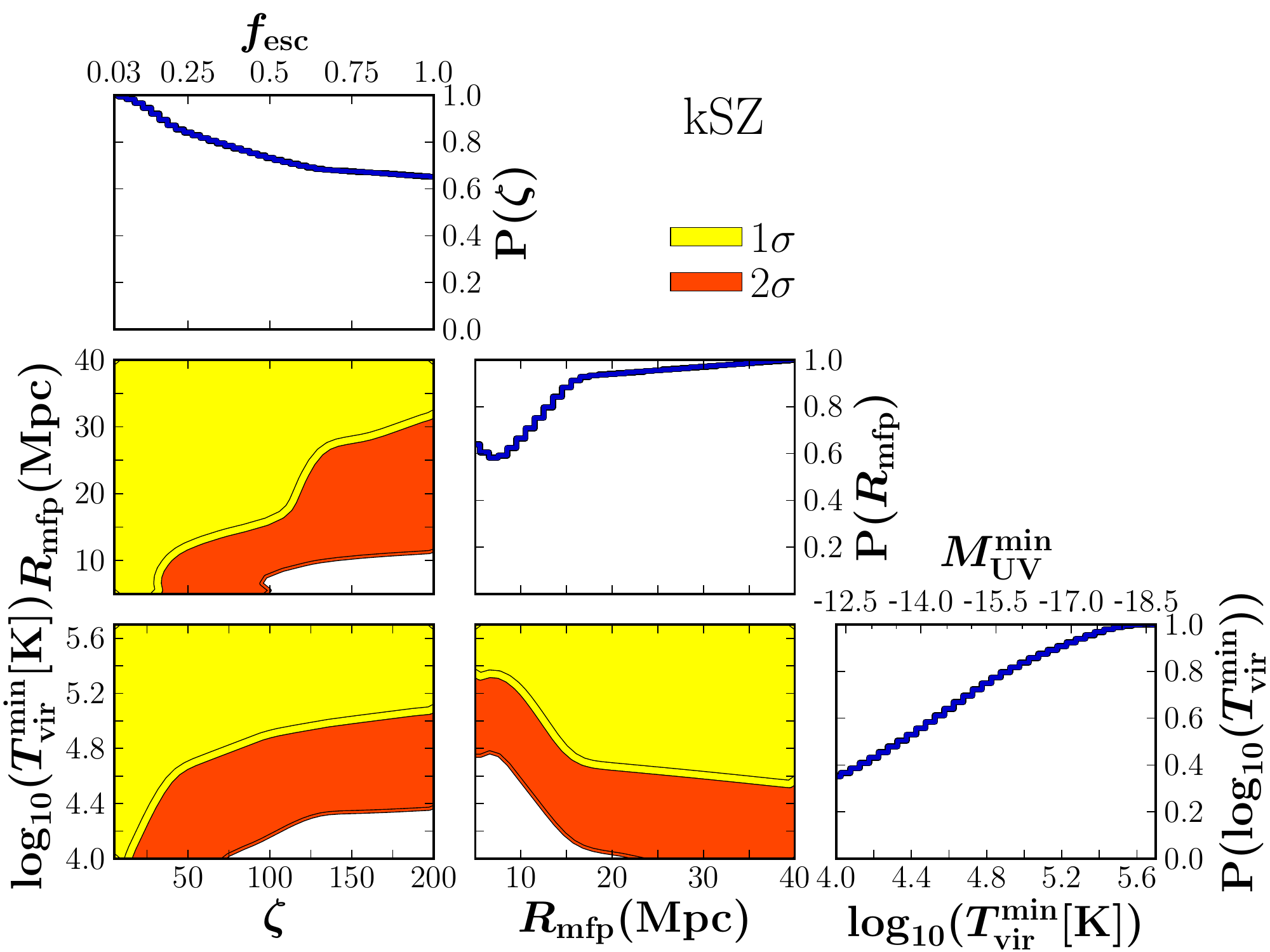}
\includegraphics[trim = 1.1cm 0.7cm 0cm 0cm, scale = 0.445]{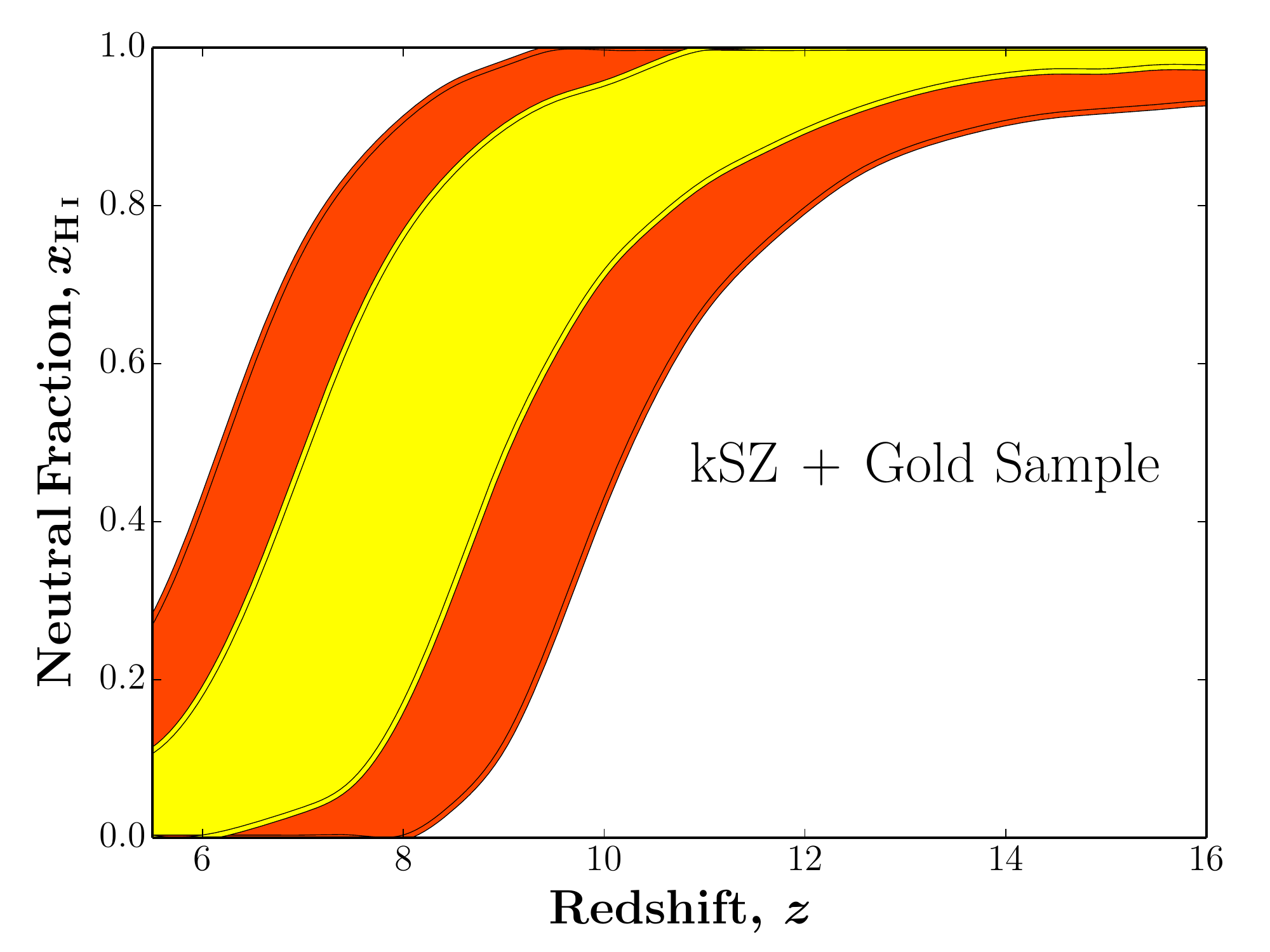}
\includegraphics[trim = 0.8cm 0.2cm 0cm 0cm, scale = 0.455]{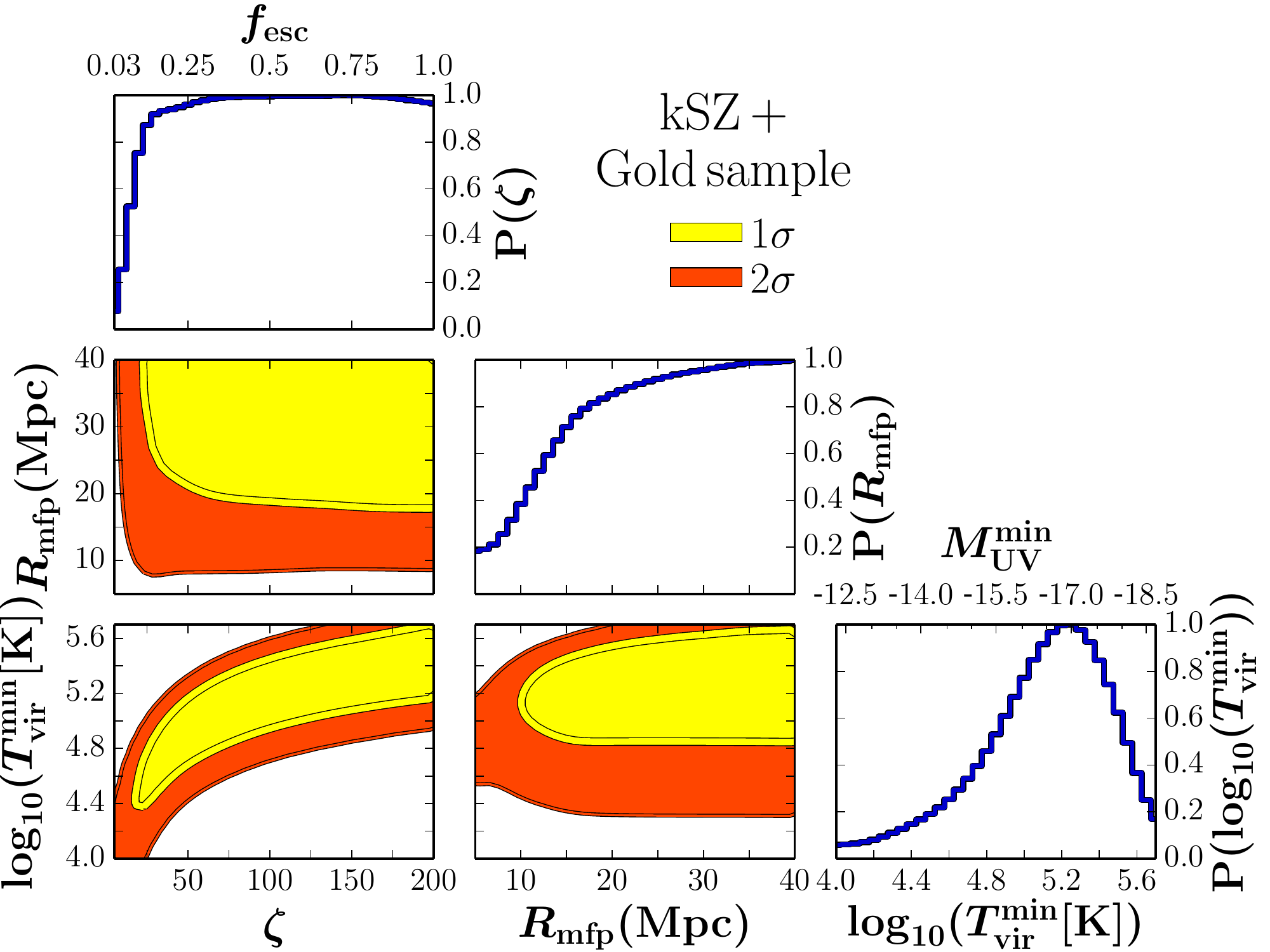}
}
\caption{
Same as Fig. \ref{fig:LAf}, but adopting instead a prior on the patchy kSZ power spectrum amplitude at $l=3000$ of $\PkSZ \approx 0.9 \pm 1.3 ~(1\sigma)~ \muKK$, motivated by the recent observations with SPT \citep{George15}.
}
\label{fig:kSZ}
\vspace{-0.5\baselineskip}
\end{figure*}

The kinetic Sunyaev-Zel'dovich (kSZ) is a secondary CMB anisotropy, sourced by photons scattering off of free elections with bulk flows.  The scattered photons either gain or lose energy, depending on the sign of the radial component of the flow.  As all CMB measurements, the total signal depends on the integral out to the LSS.  Roughly half of the expected kSZ signal is sourced by the post-reionisation IGM.  During inhomogeneous reionisation however, order unity fluctuations in the ionised fraction source a roughly equal contribution to the total kSZ, generally called the patchy kSZ.  The patchy kSZ power spectrum depends on the timing, duration and topology of reionisation.

Current efforts with the Atacama Cosmology Telescope (ACT)\footnote{http://www.physics.princeton.edu/act/} and the South Pole Telescope (SPT)\footnote{http://pole.uchicago.edu/} are measuring the amplitude of the kSZ power spectrum at a multipole of $l=3000$.  Recently, the SPT reported a detection of the total kSZ power, ${\it l}^2/[2\pi] \,C^{\rm kSZ}_{3000} = 2.9 \pm 1.3 ~\muKK$ \citep{George15}.  Subtracting out the expected contribution of $\approx 2 \muKK$ from the post-reionisation, $z \lsim 5.5$ IGM (e.g. \citealt{TBO11, SRN12, MMS12}), results in a patchy kSZ signal of $\PkSZ \approx 0.9 ~ \muKK$.  This value is on the low end of current theoretical estimates, which generally range from $\PkSZ\approx$1--3 $\muKK$ \citep{MMS12, Battaglia13}, although the observational uncertainty is still large.

Modelling the patchy kSZ\footnote{As was pointed out in \citet{MMS12} and \citet{Park13}, the same kSZ power can result from models with different astrophysics.  Hence, in this work we compute the kSZ power directly from a grid of 3D simulations, rather than adopting a template power spectrum for the signal and empirically scaling it with the duration and midpoint of reionisation.} requires larger simulation boxes than we use for our fiducial EoR models (e.g. \citealt{Iliev07kSZ}), as well as finer redshift sampling to capture the integrated signal.  Hence, we run a set of 500 Mpc, $450^3$ boxes on coarser $(\zeta, \Tvirmin, \mfp)$ grid (following \citealt{MMS12}), and then interpolate the resulting values of $\PkSZ$ to our finer grid.  We adopt a prior on the patchy kSZ power spectrum amplitude of $\PkSZ \approx 0.9 \pm 1.3 ~(1\sigma)~ \muKK$, conservatively using the full error on the total kSZ \citep{George15}.

The resulting constraints are shown in Fig. \ref{fig:kSZ}.  The relatively low observed signal favours EoR models which (i) finish late; (ii) are rapid; and/or (iii) have ionisation structure on either larger or smaller scales than $l=3000$ (for reference, this multipole corresponds to $\sim$20 Mpc at high redshifts). There is one notable difference between the kSZ constraints and those from other observations: the non-monotonic PDF for $\mfp$. Small values of $\mfp$ tend to extend reionisation, increasing the amplitude of the patchy kSZ power.  Hence these models are disfavoured by the small measured value of kSZ.  However, when the mean free path is small, the cosmic \hii{} regions tend to have characteristic diameters of $2\mfp$, producing a  peak in the shape of the patchy kSZ power spectrum around the corresponding multipoles \citep{MMS12}.  Thus models with $2\mfp < 20$ Mpc have a positive {\it slope} at $l=3000$, with the power spectrum shape peaking at higher multipoles.  As a result, the amplitude at $l=3000$ increases with decreasing $\mfp$, when $\mfp \lsim 8$ Mpc (see \citealt{MMS12} for more details).

Combining with the Gold Sample in the bottom panels of Fig. \ref{fig:kSZ}, we see that the kSZ constraint does not contribute much.  The main difference with respect to the Gold Sample alone (Fig. \ref{fig:gold}), is that very low values of $\mfp$ are disfavoured more strongly when the kSZ priors are included.

\section{Combining all EoR observations}
\label{sec:everything}

\begin{figure*}[h]
{
\includegraphics[trim = 1.1cm 0.7cm 0cm 0.3cm, scale = 0.7]{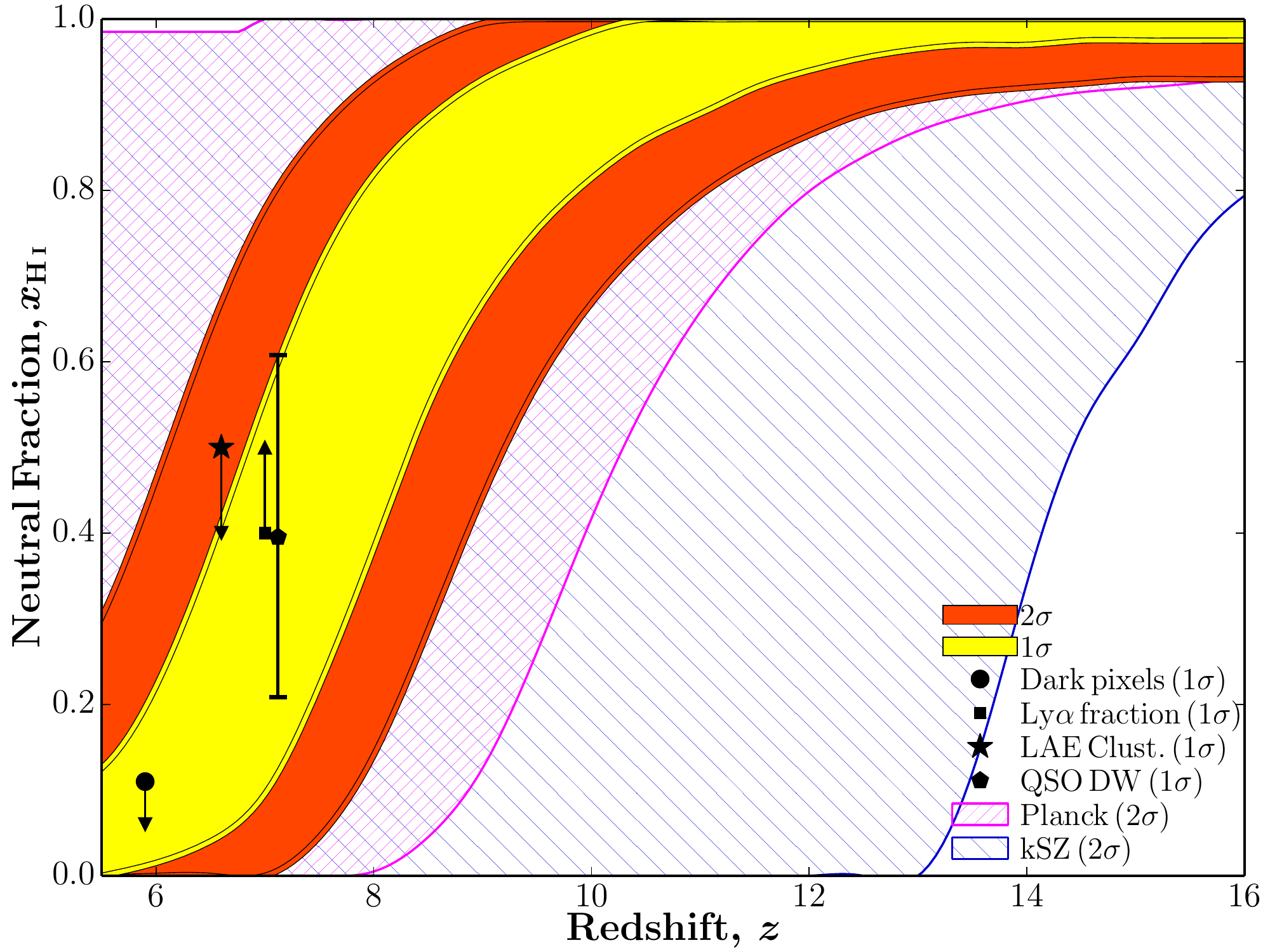}
}
\vspace{+1\baselineskip}
\caption{
Constraints on the EoR history, including all of the above-mentioned observational priors: (i) the dark fraction \citep{MMO15}, (ii) CMB optical depth \citep{Planck16}; (iii) \lya\ fraction evolution \citep{Mesinger15}; (iv) LAE clustering \citep{Ouchi10, SM15}; (v) damping wing in ULAS J1120+0641 \citep{Greig16}; (vi) patchy kSZ \citep{George15}. In order to aid the visual representation of the respective constraints we present the $2\sigma$ contours for the CMB optical depth and the patchy kSZ signal as the boundaries of the $1\sigma$ limits are masked by the constraints from the combined EoR history (c.f.\ Figs.~\ref{fig:planck} and~\ref{fig:kSZ}). For all other observations we present the $1\sigma$ limits and constraints.
} 
\label{fig:everything}
\end{figure*}

\begin{figure}
{
\includegraphics[trim = 0cm 1.2cm 0cm 0.3cm, scale = 0.42]{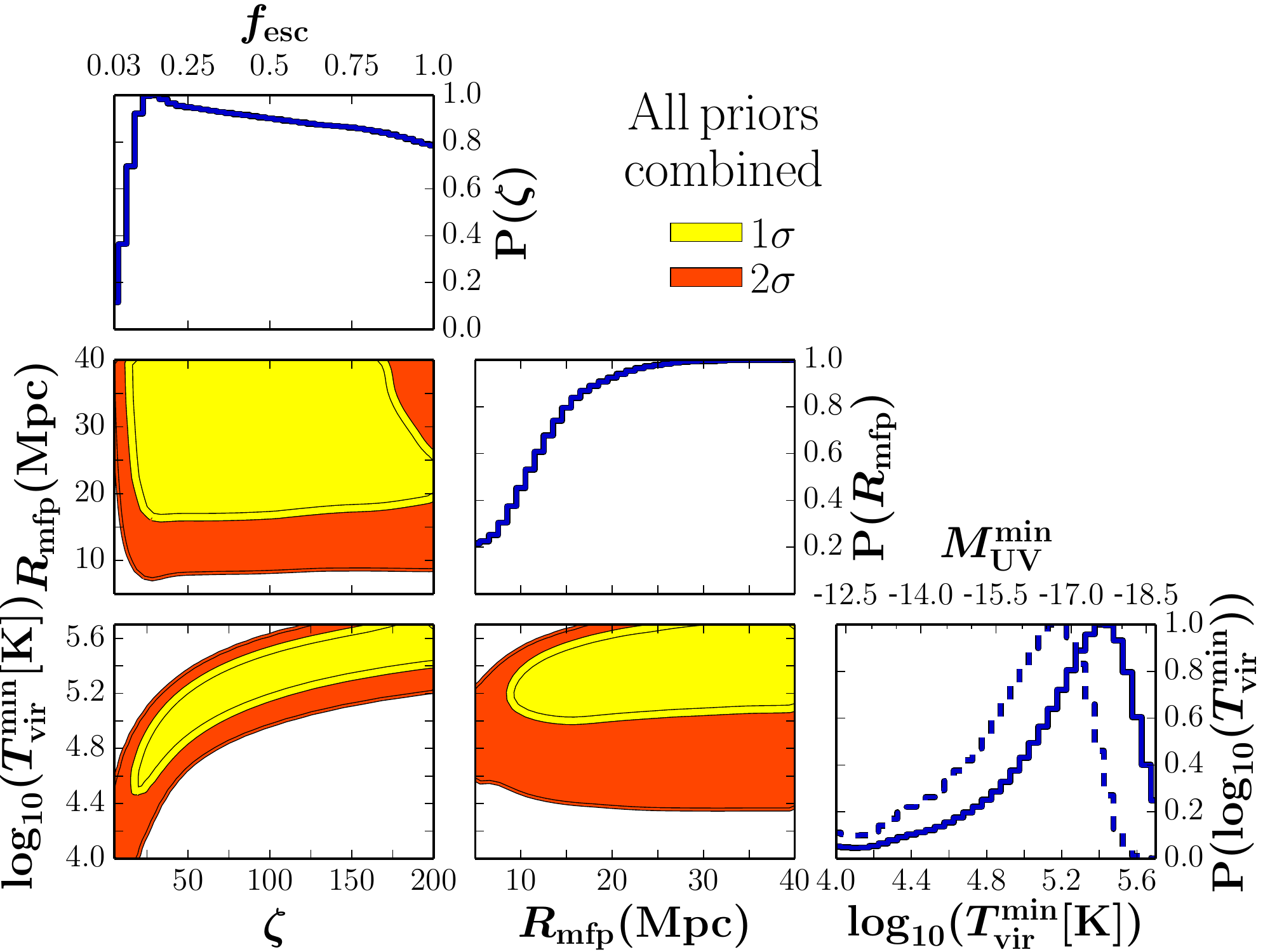}
}
\caption{
 Parameter constraints corresponding to Fig. \ref{fig:everything}.
  As in Fig. \ref{fig:planck}, the dashed blue curve in the bottom right panel shows the marginalised 1D PDF for $\Tvirmin$, but narrowing the adopted range for a flat prior on the ionising efficiency to $0<\zeta<100$ (arguably a more plausible range).
}
\label{fig:everything_params}
\vspace{-0.5\baselineskip}
\end{figure}

We now combine all of the above-mentioned observations, and show the resulting constraints on the reionisation history in Fig. \ref{fig:everything}, and the model parameters in Fig. \ref{fig:everything_params}.
At 1$\sigma$, the combined sample favours a relatively narrow range of reionisation histories, with a midpoint around $z_{\rm re} \approx=7.57^{+0.78 ~ (1\sigma)}_{-0.73 ~ (1\sigma)}$, and a duration of $\Delta_{\rm re} z \equiv z(\avenf=0.75) - z(\avenf=0.25) \approx 1.7$.  The constraints on the ionising efficiency (or alternately, the escape fraction) are weak, though models with low values of $\mfp$ and $\Tvirmin$ are disfavoured.   As discussed for the Planck priors, the $\Tvirmin$ constraints are deceptive, depending sensitively on the adopted prior over $\zeta$.

These results however should be taken with some caution, as the peak likelihood in the combined data set corresponds to $\chi^2 =2.15$.  As can be seen from these figures, and from Table 1, this mild tension is driven by the \lya\ fraction constraints.  Removing the \lya\ fraction constraints results in an EoR history which is slightly less rapid, and quantitatively very similar to the damping wing + gold sample results from Fig. \ref{fig:QSO_damp}.

\section{Model-dependent observational priors on EoR model parameters}
\label{sec:LFobs}

In the previous sections, we saw that current EoR observations, although capable of constraining the reionisation history to within $\Delta z\sim$ 1--2, are insufficient to strongly discriminate between the EoR astrophysical parameters themselves (see also \citealt{MCF15}).  This is mostly driven by the strong degeneracy between $\Tvirmin$ and $\zeta$, both of which play a crucial role in the timing of the EoR.  This will change with the advent of 21cm interferometry with second generation instruments such as HERA\footnote{http://reionization.org} and SKA\footnote{https://www.skatelescope.org}.  The reionisation morphology accessible with next-generation interferometers can provide percent-level constraints on the EoR parameters \citep{GM15}.

HERA and SKA should start taking data in the next several years, and will need a few years of observations to obtain sufficient signal to noise for precision astrophysical cosmology.  In the interim period, some model-dependent insights into EoR astrophysics can be obtained through observations of galaxies and the post-reionisation IGM. Unlike the more direct observations of $\avenf(z)$ discussed above, these observations require many assumptions to be translated into EoR parameters constraints.  For LBG LFs, these assumptions include: (i) the escape fraction $f_{\rm esc}(M_{\rm UV}, z)$; (ii) the unobserved faint-end of the LFs, whose integral (weighted by $f_{\rm esc}(M_{\rm UV}, z)$) likely dominates the ionising photon budget; and (iii) the intrinsic galactic SEDs which maps observations at $\approx1500$\AA\ to the intrinsic spectrum at $<912$\AA.

In this section, we give an example of such empirical priors on the EoR model parameters themselves.
Using the {\it Hubble Space Telescope} imaging of three Frontier Fields clusters, \citet{Atek15} recently published constraints on the faint-end of the $z\sim7$ LBG LF (see also the subsequent work by \citealt{LFL16}).  These studies find that the LBG LF remains steeply increasing down to $M_{\rm UV}\sim$-15.25. Using the fiducial scalings from Section~\ref{sec:empirical}, a conservatively-bright
choice of $M_{\rm UV}^{\rm min} \sim -16$ roughly corresponds to $\Tvirmin \sim 10^5$ K\footnote{The simple linear $\Tvir \leftrightarrow M_{\rm UV}$ mapping from Section~\ref{sec:empirical} is somewhat too steep to match the faint-end of the $z\sim7$ LF (e.g. \citealt{Finkelstein15, MOL15}).  In the context of our three-parameter EoR model, a flatter LF slope would imply either: (i) a somewhat higher value of $\Tvirmin$, for a given $M_{\rm UV}^{\rm min}$, obtained with abundance matching under the fiducial assumption of a constant duty cycle; or (ii) an ionising photon escape fraction which increases towards fainter galaxies (e.g. \citealt{PKD15}), compensating for their less-efficient star formation.  In any case, the calculation shown here is only approximate, and should be taken as a proof-of-concept for future studies when better observations of the faint end of the LF at higher redshifts, as well as insights into the scaling of $f_{\rm esc}$ with halo mass, are available.}. Requiring that $\Tvirmin \leq 10^5$ K throughout reionisation rules out large values of $\zeta$ (i.e. $f_{\rm esc}$), as reionisation would otherwise occur too early to be consistent with the {\it Planck} observations. 

\begin{figure}
{
\includegraphics[trim = 0.4cm 1.8cm 0cm 0cm, scale = 0.43]{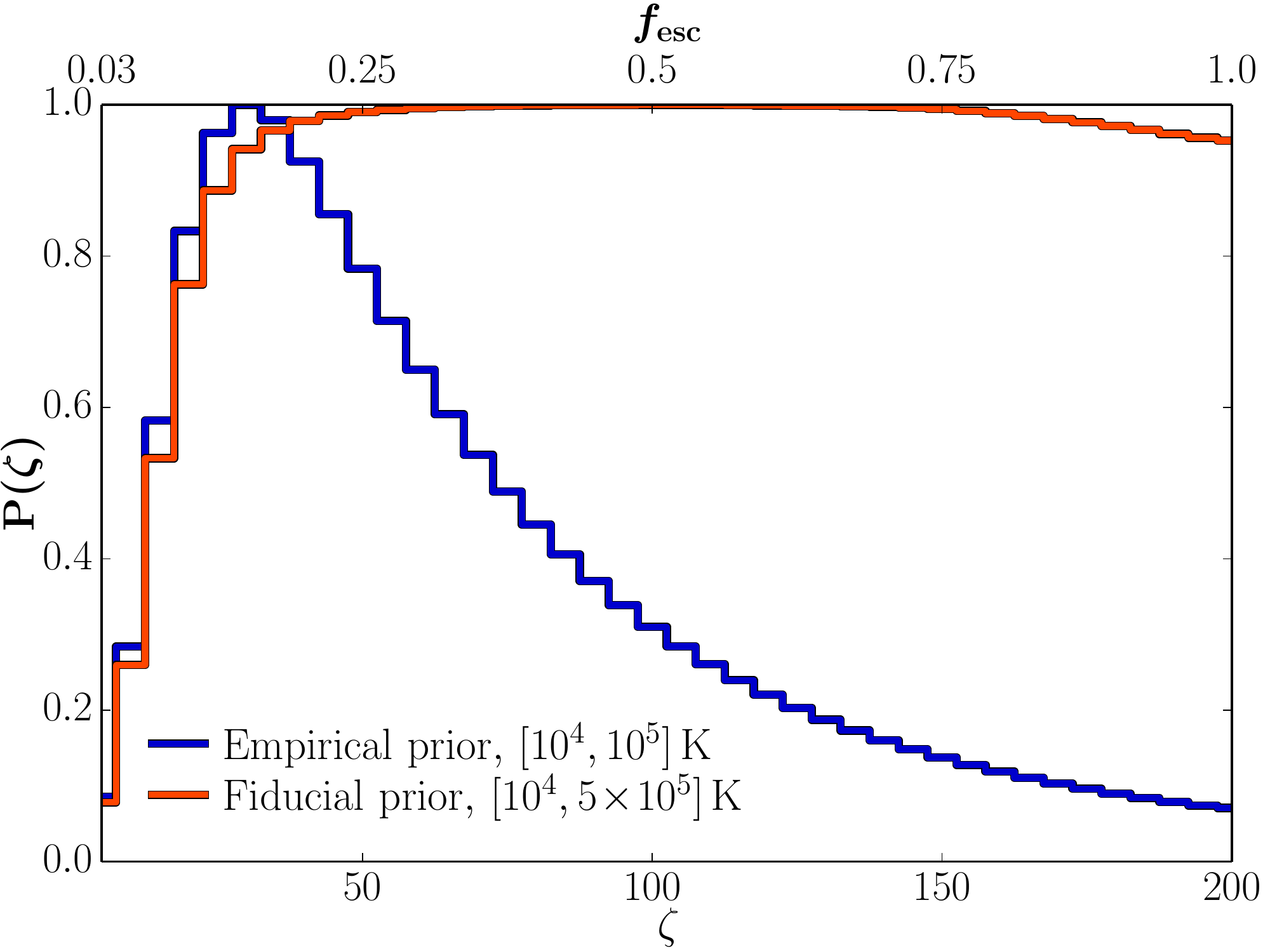}
}
\vspace{+0.5\baselineskip}
\caption{
Marginalised 1D PDFs for $\zeta$ / $f_{\rm esc}$ (top / bottom axis) obtained from the Gold Sample.  The red curve corresponds to our fiducial flat prior over $10^4 < \log(\Tvirmin/{\rm K})<5\times10^5$, while the blue curve corresponds to the empirically-motivated narrower range of $10^4 < \log(\Tvirmin/{\rm K})<10^5$.  This figure highlights that model-dependent priors on EoR parameters can help exclude regions of parameter space with strong degeneracies, resulting in tighter parameter constraints.
}
\label{fig:LFpriors}
\vspace{-0.5\baselineskip}
\end{figure}

In Fig. \ref{fig:LFpriors}, we illustrate how such an empirically-motivated prior on $\Tvirmin$ can impact our constraints on $\zeta$ (or $f_{\rm esc}$).  The red curve corresponds to the 1D marginalised PDF from our Gold Sample (as in Fig. \ref{fig:gold}), showing that $\zeta$ is essentially unconstrained due to its degeneracy with $\Tvirmin$.  On the other hand, the blue curve includes the more stringent, step-function prior of $\Tvirmin \leq 10^5$ K.  The constraints on the ionising efficiencies are improved considerably with the addition of the $\Tvirmin \leq 10^5$ K prior: $f_{\rm esc} = 0.14\substack{+0.26 \\ -0.09}$ (or $\zeta = 28\substack{+52 \\ -18}$). These constraints are quite consistent with a similar analysis done by \citet{Khaire16}, who obtain  $f_{\rm esc}$ of 0.14 -- 0.22.

A word of caution about this approach is in order.  As discussed above, our three-parameter model serves to provide a set of functions, $\avenf(z)$, to describe the reionisation history.  As long as the space of $\avenf(z)$ functions is ``reasonably'' exhaustive, it can be directly compared with EoR observations with the resulting constraints being fairly robust (i.e. not strongly dependent on the physical interpretation of the EoR parameters themselves).
In contrast, constraints on the EoR parameters themselves are much more uncertain, relying on both: (i) the accuracy of the EoR parametrisation; and (ii) the numerous assumptions necessary to connect the faint galaxy population driving reionisation to the rare bright objects we actually observe.  The results presented in this section should therefore be interpreted as a proof-of-concept.

\section{Conclusions}
\label{sec:conc}

Using a Bayesian framework, we quantify what current observations can inform us about the reionisation history of our Universe.  We MCMC sample a popular three-parameter EoR model, consisting of: (i) the ionising efficiency of the reionising sources, $\zeta$; (ii) the minimum virial temperature hosting the bulk of the reionising sources, $\Tvirmin$; and (iii) the typical horizon for ionising photons through the ionised IGM, $\mfp$.
Although these ``effective'' model parameters average over redshift and halo mass dependence, they provide an exhaustive, physically-intuitive basis set of EoR histories.

We systematically fold-in EoR observations, highlighting their individual impact.  These include: (i) the optical depth to the CMB; (ii) the dark fraction in the Lyman $\alpha$ and $\beta$ forests; (iii) the redshift evolution of galactic \lya\ emission (so-called ``\lya\ fraction''); (iv) the clustering of \lya\ emitters; (v) the IGM damping wing imprint in QSO spectra; (vi) and the patchy kinetic Sunyaev-Zel'dovich signal.  Constraints (i) and (ii) are relatively model independent, and thus comprise our ``Gold Sample''.

We find that the Gold Sample already places fairly tight constraints on the EoR history, with the epochs corresponding to an average neutral fraction of (75, 50, 25) per cent, constrained at 1$\sigma$ to $z= (9.21\substack{+1.22 \\ -1.15}, 8.14\substack{+1.08 \\ -1.00}, 7.26\substack{+1.13 \\ -0.96})$.
Folding-in more controversial, model-dependent EoR observations [(iii--vi)], strengthens these constraints by roughly 30 per cent, with the midpoint shifting down to $z_{\rm re} \approx 7.6$ and the duration (corresponding to a 50 per cent change in the neutral fraction around the midpoint) shrinking to $\Delta z_{\rm re} \approx 1.7$.
These constraints are comparable to those obtained by other studies (e.g. \citealt{MCF15,Bouwens15,Mitra16,Khaire16,Price16}) adjusted for the lower value of $\tau_e$ \citep{Planck16}, and shifted slightly to lower redshifts, driven by priors from the \lya\ fraction and the QSO damping wing imprint in ULAS J1120+0641.  The latter currently provides the tightest constraints on the EoR history.

Unfortunately, current EoR observations cannot place tight constraints on the astrophysical parameters themselves.  However, including model-dependent priors from high-$z$ galaxy observations can help.  We illustrate how observations of the faint end of the galaxy luminosity function, such as those recently obtained at $z\sim7$, can limit the allowed parameter space, resulting in 1$\sigma$ limits of $\zeta = 28\substack{+52 \\ -18}$ (or analogously using our simple conversion, $f_{\rm esc} = 0.14\substack{+0.26 \\ -0.09}$).  This framework can easily be applied to future data sets, providing improved constraints.

\vskip+0.3in

\section*{Acknowledgments}

This project has received funding from the European Research Council (ERC) under the European Union's Horizon 2020 research and innovation program (grant agreement No 638809 -- AIDA).

\bibliographystyle{mn2e}
\bibliography{ms}

\begin{thebibliography}{101}
\expandafter\ifx\csname natexlab\endcsname\relax\def\natexlab#1{#1}\fi

\bibitem[{{Ahn} {et~al.}(2012){Ahn}, {Iliev}, {Shapiro}, {Mellema}, {Koda}, \&
  {Mao}}]{Ahn12}
{Ahn} K., {Iliev} I.~T., {Shapiro} P.~R., {Mellema} G., {Koda} J., {Mao} Y.,
  2012, \apjl, 756, L16

\bibitem[{{Ali} {et~al.}(2015)}]{Ali15}
{Ali} Z.~S., {et~al.}, 2015, ArXiv e-prints:1502.06016

\bibitem[{{Alvarez} \& {Abel}(2012)}]{AA12}
{Alvarez} M.~A., {Abel} T., 2012, \apj, 747, 126

\bibitem[{{Atek} {et~al.}(2015)}]{Atek15}
{Atek} H., {et~al.}, 2015, \apj, 814, 69

\bibitem[{{Barkana} \& {Loeb}(2001)}]{BL01}
{Barkana} R., {Loeb} A., 2001, \physrep, 349, 125

\bibitem[{{Barone-Nugent} {et~al.}(2014)}]{Barone-Nugent14}
{Barone-Nugent} R.~L., {et~al.}, 2014, \apj, 793, 17

\bibitem[{{Battaglia} {et~al.}(2013){Battaglia}, {Natarajan}, {Trac}, {Cen}, \&
  {Loeb}}]{Battaglia13}
{Battaglia} N., {Natarajan} A., {Trac} H., {Cen} R., {Loeb} A., 2013, \apj,
  776, 83

\bibitem[{{Bolton} {et~al.}(2011){Bolton}, {Haehnelt}, {Warren}, {Hewett},
  {Mortlock}, {Venemans}, {McMahon}, \& {Simpson}}]{Bolton11}
{Bolton} J.~S., {Haehnelt} M.~G., {Warren} S.~J., {Hewett} P.~C., {Mortlock}
  D.~J., {Venemans} B.~P., {McMahon} R.~G., {Simpson} C., 2011, \mnras, 416,
  L70

\bibitem[{{Bosman} \& {Becker}(2015)}]{BB15}
{Bosman} S.~E.~I., {Becker} G.~D., 2015, \mnras, 452, 1105

\bibitem[{{Bouwens} {et~al.}(2015){Bouwens}, {Illingworth}, {Oesch}, {Caruana},
  {Holwerda}, {Smit}, \& {Wilkins}}]{Bouwens15}
{Bouwens} R.~J., {Illingworth} G.~D., {Oesch} P.~A., {Caruana} J., {Holwerda}
  B., {Smit} R., {Wilkins} S., 2015, ArXiv e-prints:1503.08228

\bibitem[{Carilli {et~al.}(2010)}]{Carilli10}
Carilli C.~L., {et~al.}, 2010, \apj, 714, 834

\bibitem[{{Caruana} {et~al.}(2014){Caruana}, {Bunker}, {Wilkins}, {Stanway},
  {Lorenzoni}, {Jarvis}, \& {Ebert}}]{Caruana14}
{Caruana} J., {Bunker} A.~J., {Wilkins} S.~M., {Stanway} E.~R., {Lorenzoni} S.,
  {Jarvis} M.~J., {Ebert} H., 2014, \mnras, 443, 2831

\bibitem[{{Chornock} {et~al.}(2013){Chornock}, {Berger}, {Fox}, {Lunnan},
  {Drout}, {Fong}, {Laskar}, \& {Roth}}]{Chornock13}
{Chornock} R., {Berger} E., {Fox} D.~B., {Lunnan} R., {Drout} M.~R., {Fong}
  W.-f., {Laskar} T., {Roth} K.~C., 2013, \apj, 774, 26

\bibitem[{{Choudhury} \& {Ferrara}(2006)}]{CF06}
{Choudhury} T.~R., {Ferrara} A., 2006, \mnras, 371, L55

\bibitem[{{Choudhury} {et~al.}(2008){Choudhury}, {Ferrara}, \&
  {Gallerani}}]{CFG08}
{Choudhury} T.~R., {Ferrara} A., {Gallerani} S., 2008, \mnras, 385, L58

\bibitem[{{Choudhury} {et~al.}(2015){Choudhury}, {Puchwein}, {Haehnelt}, \&
  {Bolton}}]{Choudhury15}
{Choudhury} T.~R., {Puchwein} E., {Haehnelt} M.~G., {Bolton} J.~S., 2015,
  \mnras, 452, 261

\bibitem[{{Ciardi} {et~al.}(2006){Ciardi}, {Scannapieco}, {Stoehr}, {Ferrara},
  {Iliev}, \& {Shapiro}}]{Ciardi06}
{Ciardi} B., {Scannapieco} E., {Stoehr} F., {Ferrara} A., {Iliev} I.~T.,
  {Shapiro} P.~R., 2006, \mnras, 366, 689

\bibitem[{{Croft}(1998)}]{Croft98}
{Croft} R.~A.~C., 1998, in Eighteenth Texas Symposium on Relativistic
  Astrophysics, {A.~V.~Olinto, J.~A.~Frieman, \& D.~N.~Schramm}, ed., pp.
  664--+

\bibitem[{{Dijkstra} {et~al.}(2011){Dijkstra}, {Mesinger}, \& {Wyithe}}]{DMW11}
{Dijkstra} M., {Mesinger} A., {Wyithe} J.~S.~B., 2011, \mnras, 414, 2139

\bibitem[{{Dijkstra} {et~al.}(2014){Dijkstra}, {Wyithe}, {Haiman}, {Mesinger},
  \& {Pentericci}}]{Dijkstra14}
{Dijkstra} M., {Wyithe} S., {Haiman} Z., {Mesinger} A., {Pentericci} L., 2014,
  \mnras, 440, 3309

\bibitem[{{D'Odorico} {et~al.}(2013)}]{Odorico13}
{D'Odorico} V., {et~al.}, 2013, \mnras, 435, 1198

\bibitem[{{Emberson} {et~al.}(2013){Emberson}, {Thomas}, \& {Alvarez}}]{ETA13}
{Emberson} J.~D., {Thomas} R.~M., {Alvarez} M.~A., 2013, \apj, 763, 146

\bibitem[{Fan {et~al.}(2001)}]{Fan01}
Fan X., {et~al.}, 2001, \aj, 122, 2833

\bibitem[{{Finkelstein} {et~al.}(2012)}]{Finkelstein12}
{Finkelstein} S.~L., {et~al.}, 2012, \apj, 756, 164

\bibitem[{{Finkelstein} {et~al.}(2015)}]{Finkelstein15}
---, 2015, ArXiv e-prints:1504.00005

\bibitem[{{Furlanetto} {et~al.}(2004){Furlanetto}, {Hernquist}, \&
  {Zaldarriaga}}]{FHZ04}
{Furlanetto} S.~R., {Hernquist} L., {Zaldarriaga} M., 2004, \mnras, 354, 695

\bibitem[{{Furlanetto} \& {Mesinger}(2009)}]{FM09}
{Furlanetto} S.~R., {Mesinger} A., 2009, \mnras, 394, 1667

\bibitem[{{Furlanetto} \& {Oh}(2005)}]{FO05}
{Furlanetto} S.~R., {Oh} S.~P., 2005, \mnras, 363, 1031

\bibitem[{{Furlanetto} {et~al.}(2006){Furlanetto}, {Zaldarriaga}, \&
  {Hernquist}}]{FZH06}
{Furlanetto} S.~R., {Zaldarriaga} M., {Hernquist} L., 2006, \mnras, 365, 1012

\bibitem[{{Gallerani} {et~al.}(2006){Gallerani}, {Choudhury}, \&
  {Ferrara}}]{GCF06}
{Gallerani} S., {Choudhury} T.~R., {Ferrara} A., 2006, \mnras, 370, 1401

\bibitem[{{George} {et~al.}(2015)}]{George15}
{George} E.~M., {et~al.}, 2015, \apj, 799, 177

\bibitem[{{Gnedin}(2014)}]{Gnedin14}
{Gnedin} N.~Y., 2014, \apj, 793, 29

\bibitem[{{Greig} \& {Mesinger}(2015)}]{GM15}
{Greig} B., {Mesinger} A., 2015, \mnras, 449, 4246

\bibitem[{{Greig} {et~al.}(2016{\natexlab{a}}){Greig}, {Mesinger}, {Haiman}, \&
  {Simcoe}}]{Greig16}
{Greig} B., {Mesinger} A., {Haiman} Z., {Simcoe} R.~A., 2016{\natexlab{a}},
  ArXiv e-prints:1606.00441

\bibitem[{{Greig} {et~al.}(2016{\natexlab{b}}){Greig}, {Mesinger}, {McGreer},
  {Gallerani}, \& {Haiman}}]{Greig16Meth}
{Greig} B., {Mesinger} A., {McGreer} I.~D., {Gallerani} S., {Haiman} Z.,
  2016{\natexlab{b}}, ArXiv e-prints:1605.09388

\bibitem[{{Harker} {et~al.}(2012){Harker}, {Pritchard}, {Burns}, \&
  {Bowman}}]{Harker12}
{Harker} G.~J.~A., {Pritchard} J.~R., {Burns} J.~O., {Bowman} J.~D., 2012,
  \mnras, 419, 1070

\bibitem[{{Iliev} {et~al.}(2007{\natexlab{a}}){Iliev}, {Mellema}, {Shapiro}, \&
  {Pen}}]{Iliev07}
{Iliev} I.~T., {Mellema} G., {Shapiro} P.~R., {Pen} U.-L., 2007{\natexlab{a}},
  \mnras, 376, 534

\bibitem[{{Iliev} {et~al.}(2007{\natexlab{b}}){Iliev}, {Pen}, {Bond},
  {Mellema}, \& {Shapiro}}]{Iliev07kSZ}
{Iliev} I.~T., {Pen} U.-L., {Bond} J.~R., {Mellema} G., {Shapiro} P.~R.,
  2007{\natexlab{b}}, \apj, 660, 933

\bibitem[{Iliev {et~al.}(2006)}]{Iliev06}
Iliev I.~T., {et~al.}, 2006, \mnras, 371, 1057

\bibitem[{{Jensen} {et~al.}(2013){Jensen}, {Laursen}, {Mellema}, {Iliev},
  {Sommer-Larsen}, \& {Shapiro}}]{Jensen13}
{Jensen} H., {Laursen} P., {Mellema} G., {Iliev} I.~T., {Sommer-Larsen} J.,
  {Shapiro} P.~R., 2013, \mnras, 428, 1366

\bibitem[{{Khaire} {et~al.}(2016){Khaire}, {Srianand}, {Choudhury}, \&
  {Gaikwad}}]{Khaire16}
{Khaire} V., {Srianand} R., {Choudhury} T.~R., {Gaikwad} P., 2016, \mnras, 457,
  4051

\bibitem[{Komatsu {et~al.}(2011)}]{Komatsu11}
Komatsu E., {et~al.}, 2011, \apjs, 192, 18

\bibitem[{{Koopmans} {et~al.}(2015)}]{Koopmans15}
{Koopmans} L., {et~al.}, 2015, Advancing Astrophysics with the Square Kilometre
  Array (AASKA14), 1

\bibitem[{{Kramer} \& {Haiman}(2009)}]{KH09}
{Kramer} R.~H., {Haiman} Z., 2009, \mnras, 400, 1493

\bibitem[{{Kuhlen} \& {Faucher-Giguere}(2012)}]{KF-G12}
{Kuhlen} M., {Faucher-Giguere} C.-A., 2012, \mnras, 423, 862

\bibitem[{{Liu} {et~al.}(2015){Liu}, {Mutch}, {Angel}, {Duffy}, {Geil},
  {Poole}, {Mesinger}, \& {Wyithe}}]{Liu16}
{Liu} C., {Mutch} S.~J., {Angel} P.~W., {Duffy} A.~R., {Geil} P.~M., {Poole}
  G.~B., {Mesinger} A., {Wyithe} J.~S.~B., 2015, ArXiv e-prints:1512.00563

\bibitem[{{Livermore} {et~al.}(2016){Livermore}, {Finkelstein}, \&
  {Lotz}}]{LFL16}
{Livermore} R.~C., {Finkelstein} S.~L., {Lotz} J.~M., 2016, ArXiv
  e-prints:1604.06799

\bibitem[{{Maio} {et~al.}(2013){Maio}, {Ciardi}, \& {M{\"u}ller}}]{MCM13}
{Maio} U., {Ciardi} B., {M{\"u}ller} V., 2013, \mnras, 435, 1443

\bibitem[{{Mashian} {et~al.}(2015){Mashian}, {Oesch}, \& {Loeb}}]{MOL15}
{Mashian} N., {Oesch} P., {Loeb} A., 2015, ArXiv e-prints:1507.00999

\bibitem[{{McGreer} {et~al.}(2015){McGreer}, {Mesinger}, \&
  {D'Odorico}}]{MMO15}
{McGreer} I.~D., {Mesinger} A., {D'Odorico} V., 2015, \mnras, 447, 499

\bibitem[{{McQuinn} {et~al.}(2007{\natexlab{a}}){McQuinn}, {Hernquist},
  {Zaldarriaga}, \& {Dutta}}]{McQuinn07LAE}
{McQuinn} M., {Hernquist} L., {Zaldarriaga} M., {Dutta} S., 2007{\natexlab{a}},
  \mnras, 381, 75

\bibitem[{{McQuinn} {et~al.}(2007{\natexlab{b}}){McQuinn}, {Lidz}, {Zahn},
  {Dutta}, {Hernquist}, \& {Zaldarriaga}}]{McQuinn07}
{McQuinn} M., {Lidz} A., {Zahn} O., {Dutta} S., {Hernquist} L., {Zaldarriaga}
  M., 2007{\natexlab{b}}, \mnras, 377, 1043

\bibitem[{{McQuinn} {et~al.}(2008){McQuinn}, {Lidz}, {Zaldarriaga},
  {Hernquist}, \& {Dutta}}]{McQuinn08}
{McQuinn} M., {Lidz} A., {Zaldarriaga} M., {Hernquist} L., {Dutta} S., 2008,
  \mnras, 388, 1101

\bibitem[{{McQuinn} {et~al.}(2011){McQuinn}, {Oh}, \&
  {Faucher-Gigu{\`e}re}}]{MOF11}
{McQuinn} M., {Oh} S.~P., {Faucher-Gigu{\`e}re} C.-A., 2011, \apj, 743, 82

\bibitem[{{Mesinger}(2010)}]{Mesinger10}
{Mesinger} A., 2010, \mnras, 407, 1328

\bibitem[{{Mesinger} {et~al.}(2015){Mesinger}, {Aykutalp}, {Vanzella},
  {Pentericci}, {Ferrara}, \& {Dijkstra}}]{Mesinger15}
{Mesinger} A., {Aykutalp} A., {Vanzella} E., {Pentericci} L., {Ferrara} A.,
  {Dijkstra} M., 2015, \mnras, 446, 566

\bibitem[{{Mesinger} \& {Dijkstra}(2008)}]{MD08}
{Mesinger} A., {Dijkstra} M., 2008, \mnras, 390, 1071

\bibitem[{{Mesinger} {et~al.}(2013){Mesinger}, {Ferrara}, \& {Spiegel}}]{MFS13}
{Mesinger} A., {Ferrara} A., {Spiegel} D.~S., 2013, \mnras, 431, 621

\bibitem[{{Mesinger} \& {Furlanetto}(2007)}]{MF07}
{Mesinger} A., {Furlanetto} S., 2007, \apj, 669, 663

\bibitem[{{Mesinger} {et~al.}(2011){Mesinger}, {Furlanetto}, \& {Cen}}]{MFC11}
{Mesinger} A., {Furlanetto} S., {Cen} R., 2011, \mnras, 411, 955

\bibitem[{{Mesinger} \& {Furlanetto}(2008{\natexlab{a}})}]{MF08damp}
{Mesinger} A., {Furlanetto} S.~R., 2008{\natexlab{a}}, \mnras, 385, 1348

\bibitem[{{Mesinger} \& {Furlanetto}(2008{\natexlab{b}})}]{MF08LAE}
---, 2008{\natexlab{b}}, \mnras, 386, 1990

\bibitem[{{Mesinger} {et~al.}(2016){Mesinger}, {Greig}, \&
  {Sobacchi}}]{MGS2016}
{Mesinger} A., {Greig} B., {Sobacchi} E., 2016, \mnras, 459, 2342

\bibitem[{{Mesinger} \& {Haiman}(2004)}]{MH04}
{Mesinger} A., {Haiman} Z., 2004, \apjl, 611, L69

\bibitem[{{Mesinger} \& {Haiman}(2007)}]{MH07}
---, 2007, \apj, 660, 923

\bibitem[{{Mesinger} {et~al.}(2004){Mesinger}, {Haiman}, \& {Cen}}]{MHC04}
{Mesinger} A., {Haiman} Z., {Cen} R., 2004, \apj, 613, 23

\bibitem[{{Mesinger} {et~al.}(2012){Mesinger}, {McQuinn}, \& {Spergel}}]{MMS12}
{Mesinger} A., {McQuinn} M., {Spergel} D.~N., 2012, \mnras, 422, 1403

\bibitem[{{Mitra} {et~al.}(2011){Mitra}, {Choudhury}, \& {Ferrara}}]{MCF11}
{Mitra} S., {Choudhury} T.~R., {Ferrara} A., 2011, \mnras, 413, 1569

\bibitem[{{Mitra} {et~al.}(2015){Mitra}, {Choudhury}, \& {Ferrara}}]{MCF15}
---, 2015, \mnras, 454, L76

\bibitem[{{Mitra} {et~al.}(2016){Mitra}, {Choudhury}, \& {Ferrara}}]{Mitra16}
---, 2016, ArXiv e-prints:1606.02719

\bibitem[{Mortlock {et~al.}(2011)}]{Mortlock11}
Mortlock D.~J., {et~al.}, 2011, \nat, 474, 616

\bibitem[{{Ono} {et~al.}(2012)}]{Ono12}
{Ono} Y., {et~al.}, 2012, \apj, 744, 83

\bibitem[{{Ouchi} {et~al.}(2010)}]{Ouchi10}
{Ouchi} M., {et~al.}, 2010, \apj, 723, 869

\bibitem[{{Paardekooper} {et~al.}(2015){Paardekooper}, {Khochfar}, \& {Dalla
  Vecchia}}]{PKD15}
{Paardekooper} J.-P., {Khochfar} S., {Dalla Vecchia} C., 2015, \mnras, 451,
  2544

\bibitem[{{Park} {et~al.}(2013){Park}, {Shapiro}, {Komatsu}, {Iliev}, {Ahn}, \&
  {Mellema}}]{Park13}
{Park} H., {Shapiro} P.~R., {Komatsu} E., {Iliev} I.~T., {Ahn} K., {Mellema}
  G., 2013, \apj, 769, 93

\bibitem[{{Parsons} {et~al.}(2014)}]{Parsons14}
{Parsons} A.~R., {et~al.}, 2014, \apj, 788, 106

\bibitem[{{Patil} {et~al.}(2014)}]{Patil14}
{Patil} A.~H., {et~al.}, 2014, \mnras, 443, 1113

\bibitem[{{Pentericci} {et~al.}(2011)}]{Pentericci11}
{Pentericci} L., {et~al.}, 2011, \apj, 743, 132

\bibitem[{{Planck Collaboration XIII}(2015)}]{Planck15}
{Planck Collaboration XIII}, 2015, ArXiv e-prints:1502.01589

\bibitem[{{Planck Collaboration XLVII}(2016)}]{Planck16}
{Planck Collaboration XLVII}, 2016, ArXiv e-prints:1605.03507

\bibitem[{{Pober} {et~al.}(2014)}]{Pober14}
{Pober} J.~C., {et~al.}, 2014, \apj, 782, 66

\bibitem[{{Price} {et~al.}(2016){Price}, {Trac}, \& {Cen}}]{Price16}
{Price} L.~C., {Trac} H., {Cen} R., 2016, ArXiv e-prints:1605.03970

\bibitem[{{Prochaska} {et~al.}(2010){Prochaska}, {O'Meara}, \&
  {Worseck}}]{POW10}
{Prochaska} J.~X., {O'Meara} J.~M., {Worseck} G., 2010, \apj, 718, 392

\bibitem[{{Ricotti} \& {Ostriker}(2004)}]{RO4}
{Ricotti} M., {Ostriker} J.~P., 2004, \mnras, 352, 547

\bibitem[{{Robertson} {et~al.}(2015){Robertson}, {Ellis}, {Furlanetto}, \&
  {Dunlop}}]{Robertson15}
{Robertson} B.~E., {Ellis} R.~S., {Furlanetto} S.~R., {Dunlop} J.~S., 2015,
  \apjl, 802, L19

\bibitem[{{Robertson} {et~al.}(2013)}]{Robertson13}
{Robertson} B.~E., {et~al.}, 2013, \apj, 768, 71

\bibitem[{{Schenker} {et~al.}(2014){Schenker}, {Ellis}, {Konidaris}, \&
  {Stark}}]{Schenker14}
{Schenker} M.~A., {Ellis} R.~S., {Konidaris} N.~P., {Stark} D.~P., 2014, \apj,
  795, 20

\bibitem[{{Schroeder} {et~al.}(2013){Schroeder}, {Mesinger}, \&
  {Haiman}}]{SMH13}
{Schroeder} J., {Mesinger} A., {Haiman} Z., 2013, \mnras, 428, 3058

\bibitem[{{Shaw} {et~al.}(2012){Shaw}, {Rudd}, \& {Nagai}}]{SRN12}
{Shaw} L.~D., {Rudd} D.~H., {Nagai} D., 2012, \apj, 756, 15

\bibitem[{{Simcoe} {et~al.}(2012){Simcoe}, {Sullivan}, {Cooksey}, {Kao},
  {Matejek}, \& {Burgasser}}]{Simcoe12}
{Simcoe} R.~A., {Sullivan} P.~W., {Cooksey} K.~L., {Kao} M.~M., {Matejek}
  M.~S., {Burgasser} A.~J., 2012, \nat, 492, 79

\bibitem[{{Sirko}(2005)}]{Sirko05}
{Sirko} E., 2005, \apj, 634, 728

\bibitem[{{Sobacchi} \& {Mesinger}(2014)}]{SM14}
{Sobacchi} E., {Mesinger} A., 2014, \mnras, 440, 1662

\bibitem[{{Sobacchi} \& {Mesinger}(2015)}]{SM15}
---, 2015, ArXiv e-prints:1505.02787

\bibitem[{{Stark} {et~al.}(2010){Stark}, {Ellis}, {Chiu}, {Ouchi}, \&
  {Bunker}}]{Stark10}
{Stark} D.~P., {Ellis} R.~S., {Chiu} K., {Ouchi} M., {Bunker} A., 2010, \mnras,
  408, 1628

\bibitem[{{Totani} {et~al.}(2006){Totani}, {Kawai}, {Kosugi}, {Aoki}, {Yamada},
  {Iye}, {Ohta}, \& {Hattori}}]{Totani06}
{Totani} T., {Kawai} N., {Kosugi} G., {Aoki} K., {Yamada} T., {Iye} M., {Ohta}
  K., {Hattori} T., 2006, \pasj, 58, 485

\bibitem[{{Trac} {et~al.}(2011){Trac}, {Bode}, \& {Ostriker}}]{TBO11}
{Trac} H., {Bode} P., {Ostriker} J.~P., 2011, \apj, 727, 94

\bibitem[{{Trac} \& {Gnedin}(2011)}]{TG11}
{Trac} H.~Y., {Gnedin} N.~Y., 2011, Advanced Science Letters, 4, 228

\bibitem[{{White} {et~al.}(2003){White}, {Becker}, {Fan}, \&
  {Strauss}}]{White03}
{White} R.~L., {Becker} R.~H., {Fan} X., {Strauss} M.~A., 2003, \aj, 126, 1

\bibitem[{{Yue} {et~al.}(2016){Yue}, {Ferrara}, \& {Xu}}]{YFX16}
{Yue} B., {Ferrara} A., {Xu} Y., 2016, ArXiv e-prints:1604.01314

\bibitem[{{Zahn} {et~al.}(2007){Zahn}, {Lidz}, {McQuinn}, {Dutta}, {Hernquist},
  {Zaldarriaga}, \& {Furlanetto}}]{Zahn07}
{Zahn} O., {Lidz} A., {McQuinn} M., {Dutta} S., {Hernquist} L., {Zaldarriaga}
  M., {Furlanetto} S.~R., 2007, \apj, 654, 12

\bibitem[{{Zahn} {et~al.}(2012)}]{Zahn12}
{Zahn} O., {et~al.}, 2012, \apj, 756, 65

\end{thebibliography}

\end{document}